\documentclass[]{tJOT2e}
\usepackage{flushend}
\usepackage{graphicx}
\usepackage{listings}
\usepackage{amsmath}
\usepackage{setspace}
\usepackage{color}
\usepackage{cancel}
\allowdisplaybreaks

\newcommand{\Ret}{\textit{Re}_\tau}  

\newcommand{\eg}{e.g.\,}
\newcommand{\ie}{i.e.\,}

\newcommand{\cL}{\mathbf{\cal L}}
\newcommand{\cH}{\mathbf{\cal H}}

\newcommand{\ub}{\mathbf{u}}
\newcommand{\kb}{\mathbf{k}}

\newcommand{\uk}{\mathbf{u_k}}
\newcommand{\pk}{p_\mathbf{k}}

\newcommand{\fk}{\mathbf{f_k}}
\newcommand{\sk}{\sigma_{\mathbf{k}}}
\newcommand{\sko}{\sigma_{\mathbf{k}0}}
\newcommand{\skc}{\sigma_{\mathbf{k}c}}

\newcommand{\lxp}{\lambda^+_x}
\newcommand{\lzp}{\lambda^+_z}

\newcommand{\kx}{\kappa_x}
\newcommand{\kz}{\kappa_z}
\newcommand{\om}{\omega}
\newcommand{\cp}{c^+}

\newcommand{\p}{\partial}

\begin{document}
\doi{10.1080/14685248.YYYYxxxxxx}
 \issn{1468-5248}
 \jvol{00} \jnum{00} \jyear{2016}

\markboth{M. Luhar, A.S. Sharma, and B.J. McKeon}{Journal of Turbulence}


\title{On the design of optimal compliant walls for turbulence control}

\author{M. Luhar$^{\rm a}$$^{\ast}$\thanks{$^\ast$Corresponding author. Email: luhar@usc.edu\vspace{6pt}}, 
	A.S. Sharma$^{\rm b}$, and B.J. McKeon$^{\rm c}$\\\vspace{6pt}
	$^{\rm a}${\em{University of Southern California, Los Angeles, California, USA}}\\
	$^{\rm b}${\em{University of Southampton, Highfield, Southampton, UK}}\\
	$^{\rm c}${\em{California Institute of Technology, Pasadena, California, USA}}\vspace{6pt}}
		
\received{December 2015}

\maketitle

\begin{abstract}
This paper employs the theoretical framework developed by Luhar et al. (\textit{J. Fluid Mech.}, \textbf{768}, 415-441) to consider the design of compliant walls for turbulent skin friction reduction.  Specifically, the effects of simple spring-damper walls are contrasted with the effects of more complex walls incorporating tension, stiffness and anisotropy.  In addition, varying mass ratios are tested to provide insight into differences between aerodynamic and hydrodynamic applications.  Despite the differing physical responses, all the walls tested exhibit some important common features.  First, the effect of the walls (positive or negative) is greatest at conditions close to resonance, with sharp transitions in performance across the resonant frequency or phase speed.  Second, compliant walls are predicted to have a more pronounced effect on slower-moving structures because such structures generally have larger wall-pressure signatures.  Third, two-dimensional (spanwise constant) structures are particularly susceptible to further amplification.  These features are consistent with many previous experiments and simulations, suggesting that mitigating the rise of such two-dimensional structures is essential to designing performance-improving walls.  For instance, it is shown that further amplification of such large-scale two-dimensional structures explains why the optimal anisotropic walls identified by Fukagata et al. via DNS (\textit{J. Turb.}, \textbf{9}, 1-17) only led to drag reduction in very small domains.  The above observations are used to develop design and methodology guidelines for future research on compliant walls.


\end{abstract}

\section{Introduction}
The design of compliant surfaces for turbulent skin friction reduction has attracted significant attention since the early experiments of Kramer \cite{Kramer1961}.  However, despite many experimental \cite[\eg][]{Bushnell1977,GadelHak1984,Lee1993,Choi1997,Zhang2015} and numerical \cite[\eg][]{Endo2002,Xu2003,Fukagata2008,Kim2014} efforts, there are few definitive results.  Broadly, the direct numerical simulations (DNS) and experiments both show that \textit{softer} surfaces often give rise to energetic two-dimensional (\ie spanwise constant) wave-like motions, which can cause a substantial increase in skin friction.  \textit{Harder} surfaces appear to have little impact on the flow, although some qualitative flow visualization experiments hint at an intermittent relaminarization-like phenomenon \cite{Lee1993}.

One of the major challenges associated with developing performance-enhancing surfaces is the extent of the parameter space to be explored.  Even the simplest spring-damper walls considered in DNS depend on three independent parameters: a mass ratio, a spring constant and a scholasdamping coefficient.  The viscoelastic layers tested frequently in experiments \cite{GadelHak1984,Lee1993} depend on at least five different parameters: two elastic constants which determine the shear- and longitudinal wave speeds, the mass density, a viscous relaxation time, and the layer thickness.  Independent evaluation and optimization of these parameters in experiments would be very time-consuming and expensive, while current computational capabilities limit DNS-based compliant wall design to low Reynolds numbers or small domain sizes.  The limitations of DNS-based design of compliant walls are well illustrated by the evolutionary optimization of anisotropic compliant walls pursued by Fukagata et al. \cite{Fukagata2008}.  Specifically, these simulations showed that the best walls obtained in channel flow DNS over a small domain of length $3h$ ($h$ is the channel half-height), led to a near-$200\%$ increase in drag when the domain length was doubled to $6h$.  Further, these DNS were limited to low Reynolds numbers.  The bulk Reynolds number was $Re_B = 2U_Bh/\nu = 3300$, where $U_B$ is the bulk-averaged flow speed and $\nu$ is kinematic viscosity.

The impracticality of experimental or numerical approaches in designing performance-enhancing walls indicates the need for a computationally inexpensive theoretical framework to study turbulence-compliant wall interactions.  In an effort to address this need, Luhar et al. \cite{Luhar2015} recently extended the resolvent formulation proposed by McKeon and Sharma \cite{McKeon2010}.  Under this formulation, the turbulent flow field is expressed as a superposition of propagating velocity response modes, identified via a gain-based decomposition of the Navier-Stokes equations (NSE).  Compliant surfaces are introduced via changes in the kinematic and dynamic boundary conditions.  In particular, a complex wall admittance is used to define the relationship between the pressure and wall-normal velocity at the wall.  This change in the boundary conditions leads to a change in the gain and structure of the modes, whereby a reduction in gain is interpreted as mode suppression.

Luhar et al. \cite{Luhar2015} show that this approach predicts the amplification of the quasi two-dimensional structures observed recently in DNS \cite{Kim2014} with minimal computation.  Further, the formulation also enables an optimization of surface properties (\ie wall admittance) to suppress flow structures known to be energetic in wall turbulence.  This material-blind optimization suggests that walls with negative damping are required to suppress the near-wall (NW) cycle, identified by various researchers as essential to controlling wall turbulence \cite[\eg][]{Bushnell1977}.  However, walls with positive damping could be effective against the so-called superstructures or very-large-scale motions (VLSMs) that appear at high Reynolds number.  Unfortunately, Luhar et al. \cite{Luhar2015} show that the optimal walls identified via this procedure also have negative effects elsewhere in spectral space, with slow-moving spanwise-constant structures particularly susceptible to further amplification.

The purpose of the present paper is to build on the above findings and evaluate the effect of varying wall models in greater detail, looking closely at the sensitivity to two-dimensional structures.  While Luhar et al. \cite{Luhar2015} focused primarily on a spring-damper wall, this paper introduces the effects of tension, stiffness and anisotropy, and considers the effects of varying mass ratios to contrast aerodynamic and hydrodynamic applications.  In addition, Reynolds number effects are explored briefly, and the framework is used to provide further insight into results from the aforementioned DNS-based optimization of anisotropic compliant walls pursued by Fukagata et al. \cite{Fukagata2008}, hereafter referred to as F2008.

One of the limitations of the resolvent formulation in its present form is the requirement of a mean velocity profile in the construction of the resolvent operator.  As such, the smooth and compliant wall mean velocity profiles from F2008 are also used to evaluate the sensitivity of the resolvent-based predictions to the specific form of the mean profile.

\section{Theory}
This section provides a brief review of the resolvent formulation proposed by McKeon and Sharma \cite{McKeon2010}, the extension to account for the effects of compliant walls developed by Luhar et al. \cite{Luhar2015}, and the wall model employed in DNS by F2008 \cite{Fukagata2008}.

\subsection{Resolvent Formulation}\label{sec:theory-resolvent}
The resolvent formulation proposed by McKeon and Sharma \cite{McKeon2010} considers the full turbulent velocity field, $\ub$, to be a superposition of highly amplified velocity structures, or modes, identified via a gain-based decomposition of the Fourier-transformed Navier-Stokes equations (NSE).  For each wavenumber-frequency combination $\kb = (\kx,\kz,c=\om/\kx)$, where $\kx$ and $\kz$ are the streamwise and spanwise wavenumbers, $\om$ is the frequency and $c$ is the phase speed, the NSE are interpreted as a forcing-response system\footnote{This paper focuses on turbulent channel flows but the approach can be generalized to pipe and boundary layer flows as well.}:

\vspace{-0.5cm}
\begin{align}\label{eqNSE}
\left[\begin{array}{c} \uk \\ \pk \end{array}\right]
= 
\left( -i\omega \left[\begin{array}{cc} \mathbf{I}  &  \\  &0\\ \end{array}\right] -
\left[\begin{array}{cc} \cL_\kb & -\nabla_\kb\\ \nabla_\kb^T & 0\\ \end{array}\right] \right)^{-1}
\left[\begin{array}{cc} \mathbf{I} \\  0\\ \end{array}\right] \fk
= 
\cH_\kb \fk
\end{align}

{\noindent}The nonlinear terms are interpreted as the forcing to the system, $(\ub \cdot \nabla \ub)_\kb = \fk(y) \exp {i(\kx x + \kz z - \om t)}$, and the resolvent operator, $\cH_\kb$, maps this forcing to the velocity and pressure responses, e.g. $\mathbf{\hat{u}}_\kb = \uk(y) \exp {i(\kx x + \kz z - \om t)}$.  Here, $x$, $y$ and $z$ are the streamwise, wall-normal and spanwise coordinates, respectively, and $t$ is time.  A subscript $\kb$ denotes an individual Fourier component.  In Eq.~\ref{eqNSE}, $\nabla_\kb$ and $\nabla_\kb^T$ represent the Fourier-transformed gradient and divergence operators, and $\cL_\kb$ is the linearized Navier-Stokes operator:

\vspace{-0.5cm}
\begin{align}\label{eqLinearOperator}
\cL_\kb
&= 
\left[\begin{array}{ccc} 
-i\kx U + \Ret^{-1} \nabla_\kb^2  & -\p U/\p y & 0 \\
0 & -i\kx U + \Ret^{-1} \nabla_\kb^2 & 0 \\
0 & 0 & -i\kx U + \Ret^{-1} \nabla_\kb^2 \\
\end{array}\right],
\end{align}

{\noindent}where $U(y)$ is the mean velocity profile and $\Ret = u_\tau h / \nu$ is the friction Reynolds number.  The variables $u_\tau$, $h$, and $\nu$ represent the friction velocity, channel half-height, and kinematic viscosity, respectively.  $\nabla_\kb^2 = [-\kx^2 + \p^2/\p y^2 - \kz^2]$ is the Fourier-transformed Laplacian.

A singular value decomposition (SVD) of the discretized resolvent operator $\cH_\kb = \sum_{m}^{} \psi_m(y)\sigma_m \phi^*_m(y)$ yields a set of orthonormal forcing ($\phi_m$) and response ($\psi_m$) modes, ordered based on the input-output gain ($\sigma_1>\sigma_2>\sigma_m>...$).  Forcing in the direction of the $m^{th}$ forcing mode with unit amplitude results in a response in the direction of the $m^{th}$ response mode amplified by factor $\sigma_m$.  Thus, forcing $\fk(y)=\phi_1(y)$ creates a response $[\uk(y),\pk(y)]^T = \sigma_1 \psi_1(y)$.  Note that the resolvent operator is scaled prior to performing the SVD to enforce an $L^2$ norm for the velocity, $\uk$, and forcing, $\fk$ \cite{Luhar2015}.

In general, for $\kb$ combinations energetic in natural turbulence, the resolvent operator tends to be low rank \cite{McKeon2010,Moarref2013}.  A limited number of input directions are highly amplified, often with $\sigma_1 \gg \sigma_2$, and so the velocity and pressure fields can be reasonably approximated by the first response mode $[\uk(y),\pk(y)]^T \sim \psi_1(y)$.  Recent studies show that this rank-1 approximation captures many of the key features of wall-bounded turbulent flows, including the emergence of coherent structures and their footprint in the wall pressure field \cite{Sharma2013,Luhar2014b}.  Further, the rank-1 modes also form useful building blocks for low-order models of flow control \cite{Luhar2014a}.  As a result, the rest of this paper only considers the first singular values and modes, dropping the subscript $1$ for convenience.  Extending the analysis to consider further singular values and modes is straightforward.  

As discussed below, the effect of the compliant wall is introduced in this framework via the boundary conditions for the velocity and pressure fields.  This change in the boundary conditions modifies the mode structure and singular value relative to the rigid wall case.  A reduction in $\sigma$ is interpreted as mode suppression, which is deemed beneficial for control purposes.  Keep in mind that this approach essentially focuses on how the compliant wall modifies the linear amplification mechanisms in the flow.  The effect of the compliant walls on nonlinear interactions between modes, and the forcing generated due to these nonlinear interactions, is neglected.

The discretized resolvent operator in Eq.~\ref{eqNSE} is constructed using a spectral collocation method on Chebyshev points.  The differentiation matrices are computed using the MATLAB differentiation matrix suite developed by Weideman and Reddy \cite{Weideman2000}.  The SVD of the resolvent operator generally yields pairs of structurally similar response modes with near-identical singular values but differing symmetry along the channel centerline \cite{Moarref2013}.  To avoid any confusion arising from this mode pairing and to make the computation more efficient, the grid is restricted to $N$ points in the lower half-channel, with user-specified mode symmetry across the centerline, $y=1$ ($y$ is normalized by the channel half-height $h$).  The specific grid resolution required for convergence tends to be Reynolds number and wave speed dependent.  For the results presented in this paper, we employ $N = 100$ at $\Ret = 2000$ and $N = 200$ at $\Ret = 2\times 10^4$.  In both cases, the singular values had converged to $O(10^{-4})$.  For greater details on numerical implementation and convergence, the reader is referred to \cite{Luhar2015}.  As a rough estimate of computational expense, construction of the resolvent operator and computing the SVD takes approximately 0.1s on a single core of a laptop for each wavenumber frequency combination at $N=100$ and 0.5s at $N = 200$.

Note that construction of the linear operator $\cL_\kb$ in Eq.~\ref{eqLinearOperator}, and hence $\cH_\kb$, requires knowledge of the mean velocity profile $U(y)$.  The exact form of this mean profile is important since high amplification in the resolvent framework results from two mechanisms: (i) localization of the modes around the critical layer, $y_c$, where the mode speed matches the mean velocity $U(y_c)=c$, and (ii) energy transfer from the mean flow to the turbulence via the so-called lift-up mechanism, which depends on the interaction between mean shear and wall-normal velocity, $\propto v_\kb (\p U/\p y)$ \cite{McKeon2010,McKeon2013,Sharma2013}.  For the modeling and optimization efforts described in \S\ref{sec:theory-BC}-\ref{sec:theory-optimal}, the mean velocity profile is generated using a well-known turbulent eddy viscosity model for smooth-walled flows \cite{Reynolds1967}.  However, when comparing model predictions with simulation results from F2008 in \S\ref{sec:theory-F2008}, the mean velocity profiles from DNS are used.  Use of the DNS mean profiles enables \textit{a posteriori} analysis of model sensitivity to the assumed $U(y)$.  In other words, we evaluate how the predicted mode amplification changes when using the mean velocity profiles from compliant wall DNS relative to the smooth wall DNS.

\subsection{Boundary Conditions}\label{sec:theory-BC}
The effect of the compliant wall is introduced by changing the boundary conditions on velocity and pressure within the resolvent (Eq.~\ref{eqNSE}) before computing the SVD.  For wall displacement $\eta(x,z,t)$ constrained to be in the wall-normal ($\mathbf{e_y}$) direction, the kinematic boundary conditions at the wall, $\ub(y=\eta) = (\p \eta / \p t) \mathbf{e_y}$, can be expressed as the following Fourier-transformed, linearized Taylor series expansions:

\vspace{-0.5cm}
\begin{align}\label{eqKinematicBCu}
u_\kb(\eta) & \approx u_\kb(0) + \eta_\kb \frac{\p U}{\p y}\Big|_{0}
+& \sum\limits_{\kb = \kb_a - \kb_b} \bcancel{\eta_{\kb_a} \frac{\p u^*_{\kb_b}}{\p y}\Big|_{0}} + ...  &=& 0, \\ \label{eqKinematicBCv}
v_\kb(\eta) & \approx v_\kb(0)
+& \sum\limits_{\kb = \kb_a - \kb_b} \bcancel{\eta_{\kb_a} \frac{\p v^*_{\kb_b}}{\p y}\Big|_{0}} + ... &=& -i\om \eta_\kb, \\ \label{eqKinematicBCw}
w_\kb(\eta) & \approx w_\kb(0) 
+& \sum\limits_{\kb = \kb_a - \kb_b} \bcancel{\eta_{\kb_a} \frac{\p w^*_{\kb_b}}{\p y}\Big|_{0}} + ... &=& 0,
\end{align}

{\noindent}where $\eta_\kb$ represents the Fourier coefficient for the wall displacement at wavenumber-frequency combination $\kb = (\kx,\kx,\om)$.  The neglected quadratic terms are shown for reference.

%
%

The use of these linearized kinematic boundary conditions is one of the key limitations of the present approach since the neglected higher-order terms in Eq.~\ref{eqKinematicBCu}-\ref{eqKinematicBCw} can become important for large wall deflection.  However, retaining terms of quadratic or higher order in the fluctuations would require a coupled nonlinear model allowing for interactions between resolvent modes across all wavenumber-frequency combinations that can interact and force the Fourier mode of interest \cite{Luhar2015,Duvvuri2015triadic}, which is outside of the scope of the current effort.  Similarly, note that the nonlinear terms arising from the Fourier mode being considered would appear in the boundary conditions for higher harmonics.  For instance, the quadratic terms $(1/2)\eta_\kb^2 (\p^2 U/\p y^2)_{y=0} + \eta_\kb (\p u_\kb/\p y)_{y=0}$ would appear in the kinematic boundary condition for the streamwise velocity of the mode with wavenumber-frequency combination $(2\kx,2\kz,2\om)$.  It can be shown that the magnitude of the first term in the above expression only becomes important relative to the retained linear term, $\eta_\kb (\p U/\p y)_{y=0}$, when the wall deformation is $O(1)$.  However, the magnitude of the second term can be significant for energetic modes with near-wall gradients, $(\p u_\kb/\p y)_{y=0}$, comparable to the mean velocity gradient, $(\p U/\p y)_{y=0}$.  Thus, while these terms do not directly influence the Fourier mode being considered, they can be important elsewhere in spectral space.  Also keep in mind that the linearized boundary conditions require an estimate of the mean shear at the wall (Eq.~\ref{eqKinematicBCu}), which is assumed to correspond to the prescribed, smooth wall mean velocity profile.  This assumption breaks down if the compliant wall significantly alters the near-wall mean flow.

The dynamic boundary condition at the wall is expressed as a mechanical admittance, $Y$, linking wall-normal velocity and pressure:

\vspace{-0.5cm}
\begin{equation}\label{eqDynamicBC}
v_\kb(0) = Y p_\kb(0).
\end{equation}

{\noindent}$Y$ dictates the relative phase and amplitude of the wall-normal velocity and the pressure at the wall.  As such, it can be used to represent walls of known material properties.  For example, the most commonly used model for compliant walls involves a tensioned plate supported on a bed of springs and dampers.  For such walls, the admittance can be expressed as \cite{Xu2003}:

\vspace{-0.5cm}
\begin{equation}\label{eqAdmittance}
Y = \frac{i \omega}{-C_m \om^2 - i \om C_d + C_{ke}}
\end{equation}

{\noindent}where $C_m$ and $C_d$ are the dimensionless mass ratio and damping coefficient, and

\vspace{-0.5cm}
\begin{equation}\label{eqEffectiveSpring}
C_{ke} = C_k + C_t k^2 + C_s k^4
\end{equation}

{\noindent}is a wavenumber-dependent effective spring constant, with $k^2 = (\kx^2 + \kz^2)$.  The parameters $C_k$, $C_t$ and $C_s$ represent the dimensionless spring constant, tension and flexural rigidity.  All of the above parameters are normalized based on the channel half-height $h$, friction velocity $u_\tau$ and fluid density $\rho$.

\subsection{Optimal Walls}\label{sec:theory-optimal}
In addition to evaluating the effects of the wall parameters individually, the resolvent framework can also be used to solve the inverse problem: finding an optimal $Y$ that leads to the most favorable effect on the turbulent flow structures of interest.  Luhar et al. \cite{Luhar2015} pursued this optimization for modes resembling the NW-cycle and VLSMs at friction Reynolds number $\Ret = u_\tau h/\nu = 2000$.  The NW-cycle was represented by the wavenumber-frequency combination $\kb = (\kx,\kz,c^+) = (12,120,10)$ and the VLSMs were represented by $\kb = (1,10,16)$.  These wavenumbers translate into structures of streamwise and spanwise wavelength $(\lxp,\lzp) \approx (1050,105)$ and $(\lxp,\lzp) \approx (12500,1250)$, respectively.  Optimality was defined in two different ways:  walls that lead to the greatest mode suppression (\ie lowest $\sk$) or the largest reduction in the channel-integrated Reynolds stress contribution from the mode \cite[per][]{Fukagata2002}.  Throughout this paper, a superscript $+$ denotes normalization with respect to $u_\tau$ and $\nu$.

For brevity, this paper focuses primarily on the optimal gain-reducing wall for modes resembling the VLSMs at $\Ret=2000$.  A simple pattern search procedure shows that a wall with admittance $Y = -2.0385 - 0.4387 i$ leads to the greatest reduction in singular value for such modes, with the ratio of compliant to rigid-wall (null-case) singular values being $\skc / \sko = 0.52$.  Note that this optimization is blind to the physical properties of the compliant walls.  Designing a wall with the appropriate admittance would then become an engineering problem (or perhaps one for material scientists).  For walls characterized by Eq.~\ref{eqAdmittance}-\ref{eqEffectiveSpring}, this optimal admittance can be realized through any combination of springs, tension and stiffness.  To evaluate how these factors affect performance, particularly with respect to the excitation of spanwise-constant modes, we test the different walls listed in Table~\ref{tab:walls}, each of which has admittance $Y = -2.0385 - 0.4387 i$ for $\kb = (1,10,16)$.

\begin{table}
	\centering
	\caption{Different walls optimized to suppress resolvent modes resembling VLSMs at $Re_\tau = 2000$.  The damping coefficient is $C_d = 0.4688$ in all cases.}
	\begin{tabular}{lcccc}
	\hline
	Case  		& $C_m$  	& $C_k$  	& $C_s$  	& $C_t$ \\
	\hline
	base		& 2     	& 510.4 	& 0     	& 0 \\
	low $C_m$ 	& 0.2    	& 49.59 	& 0     	& 0 \\
	high $C_m$ 	& 20    	& 5118  	& 0     	& 0 \\
	tension 	& 2     	& 0     	& 0     	& 5.053 \\
	stiffness 	& 2     	& 0     	& 0.0500  	& 0 \\
	anisotropy	& 2     	& 0     	& 0     	& $C_{tx} = 288$\\
	& 	     	& 	     	& 	    	& $C_{tz} = 2.224$\\
	\hline
	\end{tabular}%
	\label{tab:walls}%
\end{table}%

The base case is the wall evaluated by Luhar et al. \cite{Luhar2015}, which represents a simple spring-damper system such that $C_{ke}=C_k$ and $C_m = 2$.  The high and low mass ratio cases ($C_m =20$ and $C_m=0.2$, respectively) are similar but require different spring constants to counteract the changes in $C_m$.  The next two cases in Table~\ref{tab:walls} remove the spring support but introduce the effects of tension and stiffness, such that $C_{ke} = C_t (\kx^2 + \kz^2)$ and $C_{ke} = C_s (\kx^4 + 2\kx^2 \kz^2 + \kz^4)$, respectively.  The last case introduces the effects of anisotropy through differing streamwise and spanwise tension, such that $C_{ke} = C_{tx}\kx^2 + C_{tz}\kz^2$.  Despite the physical differences, all of the walls are resonant just below the mode frequency $\om = 16$ for $\kx = 1$ and $\kz = 10$.  Specifically, the resonant frequency is $\om_r = \om_n \sqrt{1-2\zeta^2} = 15.97$, where $\om_n = \sqrt{C_{ke}/C_m}$ is the undamped natural frequency of the wall and $\zeta = C_d/(2\sqrt{C_{ke} C_m})$ is the damping factor.

One of the key advantages of the resolvent formulation is that it can be extended to higher Reynolds numbers with limited computational penalty.  Therefore, to test for Reynolds number effects, we also consider compliant walls optimized for VLSM-type modes at $\Ret = 2\times 10^4$, which we assume are characterized by wavenumber-frequency combination $\kb = (\kx,\kz,\cp) = (1,10,19)$ \cite{Marusic2010}.  For these modes, the pattern search algorithm suggests that walls with admittance $Y = -0.418 + 0.099 i$ lead to the greatest reduction in singular value, with $\skc/\sko = 0.30$.  For a spring-damper wall with mass ratio $C_m = 2$, this optimal admittance translates into stiffness $C_k (=C_{ke})= 732$ and damping $C_d = 2.26$.  The mean velocity profile for this high Reynolds number case is again obtained using an eddy viscosity formulation \cite{Reynolds1967}, while the grid resolution is increased to $N=200$ points for convergence.

\subsection{DNS-based Wall Optimization by Fukagata et al. 2008}\label{sec:theory-F2008}
\begin{figure}
	\begin{center}
		\includegraphics[width=6cm]{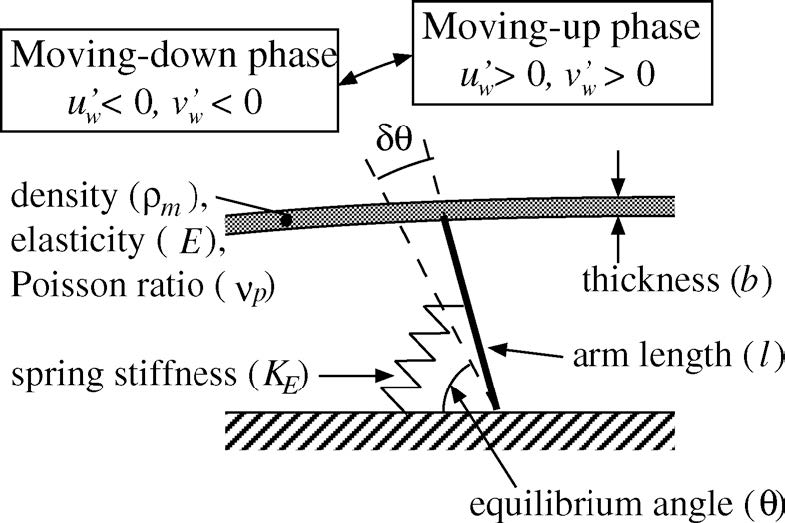}%
		\caption{Anisotropic compliant wall model employed in F2008 (image reproduced from \cite{Fukagata2008}).}
		\label{fig:F2008}
	\end{center}
\end{figure}

The anisotropic wall model employed by F2008 was introduced by Carpenter and Morris \cite{Carpenter1990anisotropic} to mimic earlier experiments performed by Grosskreutz \cite{Grosskreutz1971}.  As shown in Fig.~\ref{fig:F2008}, this wall involves an elastic plate that rests on spring-supported links that rotate about an equilibrium angle of $\theta \le \pi/2$ relative to the horizontal plane.  Similar to Eq.~\ref{eqKinematicBCu}-\ref{eqEffectiveSpring}, the boundary conditions for this wall can be expressed in terms of a single displacement variable $\eta_\kb$, which is defined as the displacement of the tip of the rigid link in this case.  Specifically, the kinematic boundary conditions are:

\vspace{-0.2cm}
\begin{equation}\label{eqKinematicBCF2008}
u_\kb(0) = -i \om \eta_\kb \sin \theta; \:\: v_\kb(0) = -i \om \eta_\kb \cos \theta; \:\: w_\kb(0) = 0,
\end{equation}

{\noindent}while the dynamic boundary condition can be expressed as:

\vspace{-0.2cm}
\begin{equation}\label{eqDynamicBCF2008}
\left[ -\rho_m b \om^2 - i \om C_d + \left\{ \frac{E b^3}{12(1-\nu_p^2)} k^4 \cos^2 \theta  + E b k^2 \sin^2 \theta + C_k \right\} \right] \eta_\kb 
= g_\kb,
\end{equation}

{\noindent}where $\rho_m$, $b$, $E$, $\nu_p$ are the dimensionless plate density, thickness, elastic modulus, and Poisson's ratio, respectively.  The forcing function on the right-hand side, $g_\kb$, is defined in Eq.~\ref{eqForcingF2008} below.  As before, $k^2 = (\kx^2 + \kz^2)$, $C_d$ is the damping coefficient, and $C_k$ is the spring stiffness.  The quantity inside the curly brackets in Eq.~\ref{eqDynamicBCF2008} can once again be considered a wavenumber-dependent effective stiffness:

\vspace{-0.2cm}
\begin{equation}\label{eqCkeF2008}
C_{ke} = \frac{E b^3}{12(1-\nu_p^2)} k^4 \cos^2 \theta  + E b k^2 \sin^2 \theta + C_k,
\end{equation}

{\noindent}leading to an undamped natural frequency of $\om_n = \sqrt{C_{ke}/(\rho_m b)}$.

Note that, unlike the compliant wall model discussed in \S\ref{sec:theory-BC}, here all variables are normalized using twice the bulk-averaged velocity, $2U_B$.  As a result, the bulk-averaged Reynolds number $Re_B = 2U_Bh/\nu = 3300$, which was kept constant in the DNS, replaces the friction Reynolds number in the linear operator (Eq.~\ref{eqLinearOperator}).  This bulk-averaged Reynolds number corresponds to $\Ret \approx 110$ ($U_B \approx 15 u_\tau$) for the smooth-wall case.  The wall is forced by a combination of pressure and turbulent stresses, such that the forcing function on the right-hand side of Eq.~\ref{eqDynamicBCF2008} is

\vspace{-0.2cm}
\begin{equation}\label{eqForcingF2008}
g_\kb = \left[ \left( -p_\kb + \frac{2}{Re_B} \frac{\p v_\kb}{\p y} \right) \cos \theta + \frac{1}{Re_B} \left( \frac{\p u_\kb}{\p y}+ i \kx v_\kb \right) \sin \theta \right]_{y=0}.
\end{equation}

Another important feature of this anisotropic wall model is the kinematic constraint imposed by the rotating link, which ensures that $u_\kb$ and $v_\kb$ are in phase at the wall (see Fig.~\ref{fig:F2008} and Eq.~\ref{eqKinematicBCF2008} above; \cite{Fukagata2008}).  As a result, the mean turbulent Reynolds stress, $-re(0.5 u_\kb^* v_\kb)$, where $re()$ denotes the real component and $()^*$ denotes a complex conjugate, is always negative at the wall.  Per the Fukagata-Iwamoto-Kasagi identity \cite{Fukagata2002}, this reduction in the Reynolds shear stress is expected to decrease momentum transfer towards the wall, leading to a reduction in skin friction.

The DNS-based evolutionary optimization of wall parameters pursued by F2008 in a small domain of length and width $3h$ suggested that compliant walls with the properties listed in Table~\ref{tab:F2008} were optimal\footnote{Per F2008, the search algorithm had not converged but the available computational time had been exhausted.}, leading to a reduction in drag of $8.3\%$ (case A1 in F2008).  However, the same wall led to a near $200\%$ \textit{increase} in drag in a domain of length $6h$.  This increase in drag was accompanied by the emergence of energetic two-dimensional wavelike structures at the wall that spanned the length of the domain (\ie modes with $\kz=0$ and $\kx = 2\pi/6$), and substantial changes in the mean velocity profile (see Fig.~\ref{fig:Fukagatakz0}a).  While domain-spanning wavelike motions were also observed in the small domain simulations, there was a near $100\%$ increase in the root-mean-square (rms) wall displacement in the large domain DNS.

In \S\ref{sec:results-F2008} below, it is shown that the resolvent framework is able to anticipate some of this deterioration in performance.  Specifically, the framework predicts that two-dimensional modes with wave numbers smaller than $\kx = 2\pi/3$ (wavelengths greater than $3h$) are susceptible to significant further amplification over compliant walls with the properties listed in Table~\ref{tab:F2008}.

\begin{table}
	\centering
	\caption{Properties of one optimal wall (case A1) identified in F2008 \cite{Fukagata2008}.  The wall thickness and Poisson' ratio were fixed at $b = 0.01$ and $\nu_p = 0.5$, respectively.  Values in the brackets below denote the initial ranges specified for the optimization.}
	\begin{tabular}{ccccc}
	\hline
	$\rho_m$ & $E$ & $C_k$ & $C_d$ & $\theta$ [deg.] \\ 
	$1.23$ & $3.00\times 10^{-3}$ & $4.25\times 10^{-5}$ & $2.21 \times 10^{-5}$ & $62.7$ \\ 
	$[0.1,10]$ & $7.2 \times 10^{[-4,2]}$ & $1.1 \times 10^{[-5,-2]}$ & ${1.0 \times 10^{[-4,-2]}}$ & $[30,90]$ \\
	\hline
	\end{tabular}%
	\label{tab:F2008}%
\end{table}%

\section{Results}

\subsection{Effect of Mass Ratio}\label{sec:results-CM}

\begin{figure}
\begin{center}
	\subfigure[]{\resizebox*{4.5cm}{!}{\includegraphics{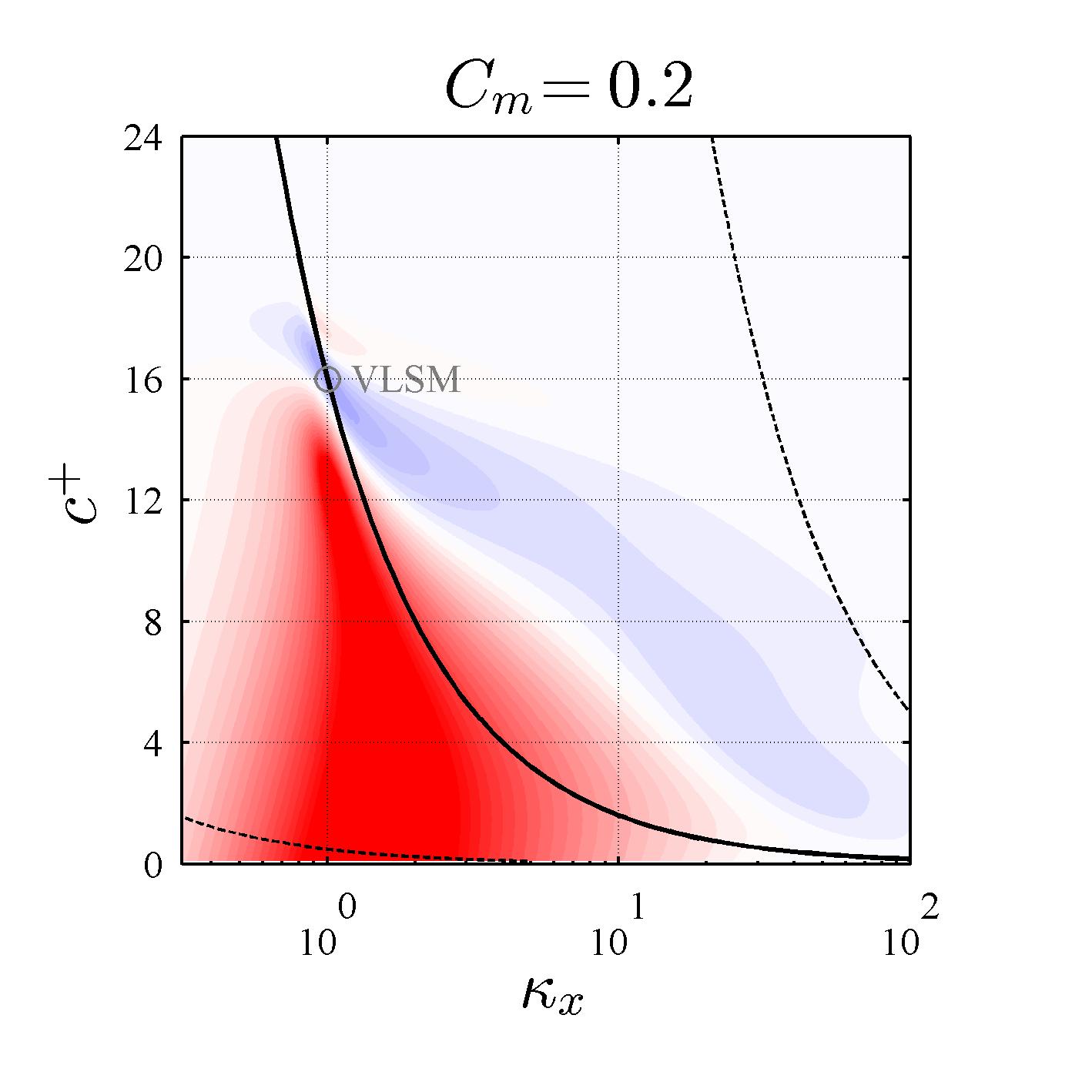}}}%
	\subfigure[]{\resizebox*{4.5cm}{!}{\includegraphics{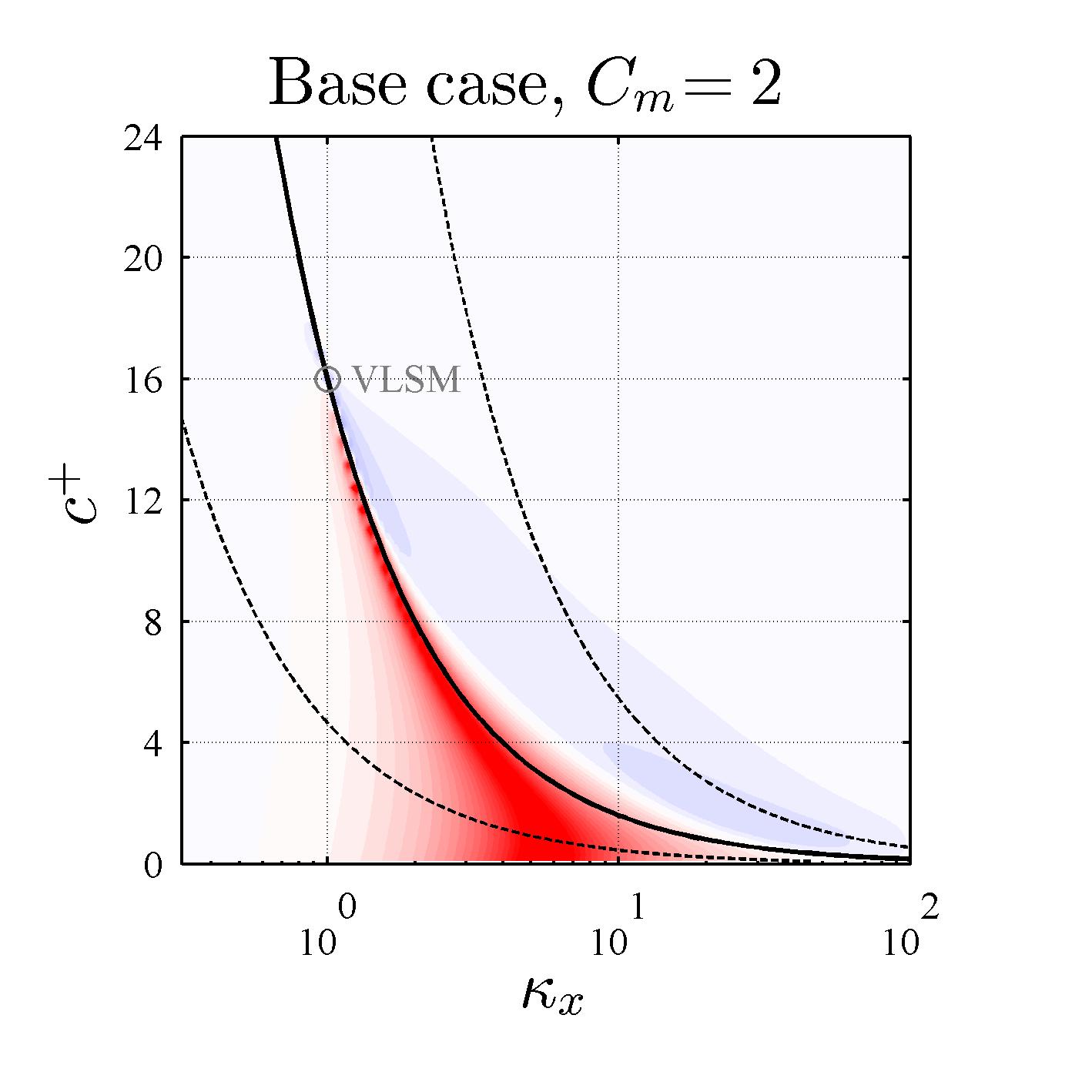}}}%
	\subfigure[]{\resizebox*{4.5cm}{!}{\includegraphics{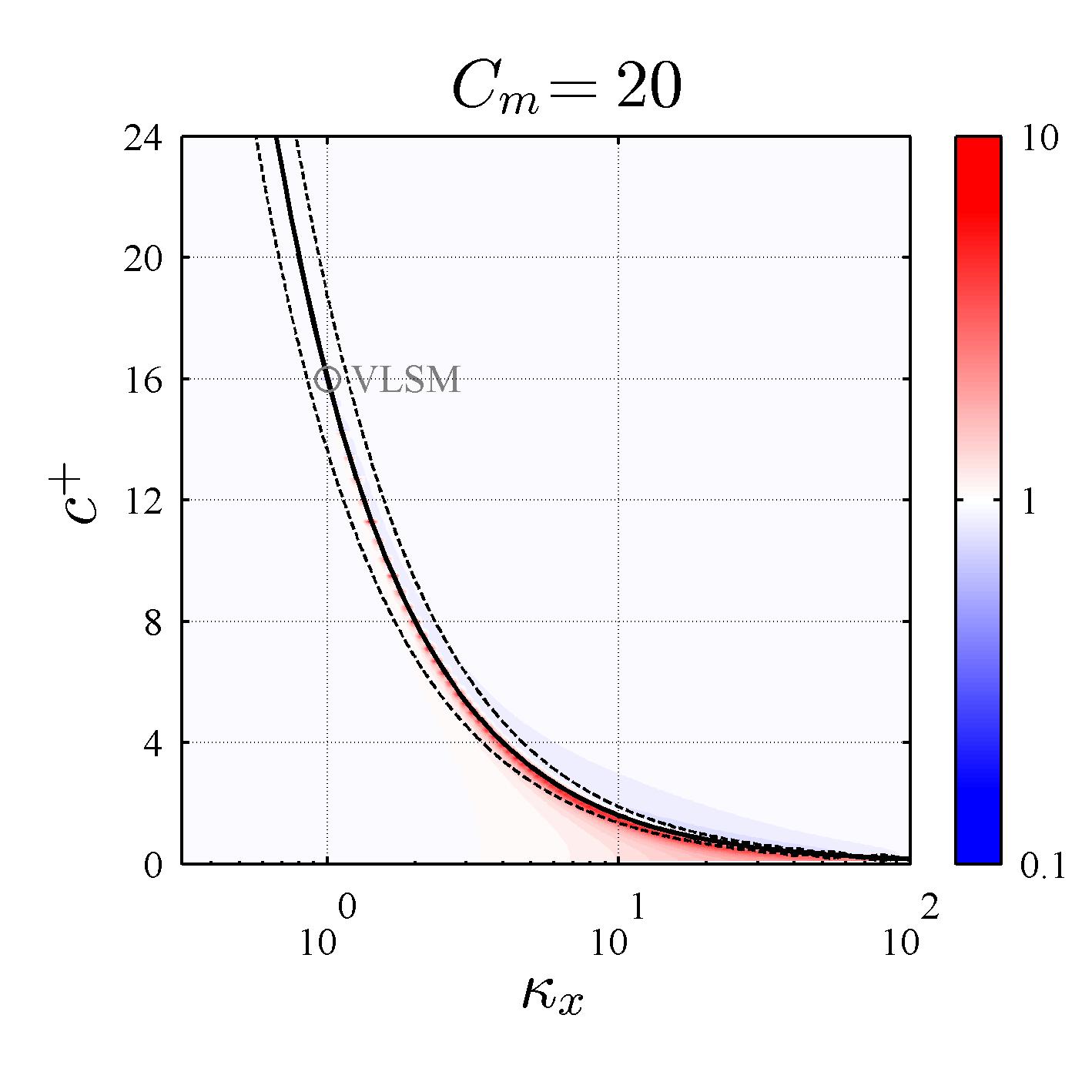}}}%
	\caption{Shaded contours showing the ratio of compliant wall to null-case singular values, $\skc/\sko$, for the low $C_m$ (a), base case (b), and high $C_m$ (c) walls listed in Table~\ref{tab:walls}.  Blue regions denote mode suppression while red regions indicate further amplification.  The solid black lines indicate the resonant frequency.  The dashed lines represent isocontours of the magnitude of the admittance $|Y|$ at level $0.01$.  All results correspond to $\kz=10$.}
	\label{fig:Cm}
\end{center}
\end{figure}

One of the key differences between aerodynamic and hydrodynamic flows over compliant walls is the mass ratio, which is determined by the ratio of the solid density to the fluid density.  While $C_m \sim O(1)$ is appropriate for hydrodynamic applications, it is expected that $C_m \sim O(10^3)$ for aerodynamic applications.  A high mass ratio translates into a much smaller wall response to fluid pressure perturbations away from resonance, which in turn means that the wall does not significantly influence the flow structures.  This is illustrated by the low $C_m$, base case, and high $C_m$ results shown in Fig.~\ref{fig:Cm}.  The spectral region over which the compliant wall has a strong influence on the singular values (positive or negative) shrinks significantly as the mass ratio is increased from $C_m = 0.2$ (Fig.~\ref{fig:Cm}a) to $C_m = 20$ (Fig.~\ref{fig:Cm}c).  As an example, for $\kx = 10$ the $C_m=0.2$ wall affects modes with speeds up to $c^+ \approx 13$, while the $C_m = 20$ wall only affects modes with speeds up to $c^+ \approx 3$.  Note that the region of influence in all cases is centered approximately around the resonant frequency (bold black line), where the magnitude of the admittance $|Y|$ peaks.

In general, the compliant walls seem to have a positive influence (suppression) on modes with frequencies higher than the resonant frequency (\ie above the solid black line) and a negative effect on modes with lower frequencies.  Although, this reverses for modes with $\kx < 1$ and $c^+ > 15$ for the low $C_m$ wall (Fig.~\ref{fig:Cm}a).  These transitions in performance may be attributed to two factors: (i) changes in the phase relationship between the pressure and velocity fields as the mode speed increases (\ie as the modes move further away from the wall), and (ii) the phase shift in the wall response, $\angle Y$, across the resonant frequency.

Also shown in Fig.~\ref{fig:Cm} are isocontours of the magnitude of the wall admittance $|Y|$ at level $0.01$ (dashed lines).  A comparison of Fig.~\ref{fig:Cm}a-c shows that the region enclosed by these isocontours reduces rapidly with increasing mass ratio.  More quantitatively, the half-power bandwidth of a spring-damper system is expected to scale as $\zeta \om_n \sim C_m^{-1}$ for $\zeta \ll 1$ \cite{CUEDMechanics}.  So the ten-fold increase in the mass-ratio translates into a roughly ten-fold decrease in the frequency bandwidth of the wall.  This bandwidth would decrease even further for $C_m \sim O(10^3)$, suggesting that compliant walls are unlikely to be practical for aerodynamic applications requiring broadband turbulence suppression without the development of novel lightweight materials.  On the other hand, the narrow bandwidth at high $C_m$ could enable more effective targeting of specific wavenumber-frequency combinations (\ie to suppress or enhance individual velocity response modes).

Note that the decrease in the spectral influence of the compliant wall is roughly consistent with the decrease in the wall bandwidth.  However, there are regions where the wall influences the flow despite low $|Y|$ (\eg for very slow modes with $c^+ < 1$) and where the wall does not have an appreciable effect even at resonance (\eg for faster modes with $c^+ > 18$).  This is because the influence of the wall is determined both by the admittance as well as the magnitude of the wall-pressure fluctuations.  In general, the magnitude of the wall pressure fields associated with the modes decreases with increasing mode speed $c^+$ \cite[\ie as the modes move further away from the wall, see][]{Luhar2014b}, and so slower modes are likely to interact with compliant walls to a larger extent.

Keep in mind that, even over compliant surfaces, there is unlikely to be significant turbulent activity very close to the wall.  As such, paying close attention to response modes with $c^+ < 4$ (\ie corresponding to the viscous sublayer under Taylor's hypothesis) has limited utility except in cases with very high amplification resulting from resonance.  Other important exceptions to this rule are the slow moving two-dimensional structures observed in previous experiments and DNS \cite{GadelHak1984,GadelHak1986,Fukagata2008,Kim2014}, which are discussed in greater detail below.

\subsection{Comparing Springs, Tension and Stiffness}\label{sec:results-KTS}
\begin{figure}
\begin{center}
	\subfigure[]{\resizebox*{4.5cm}{!}{\includegraphics{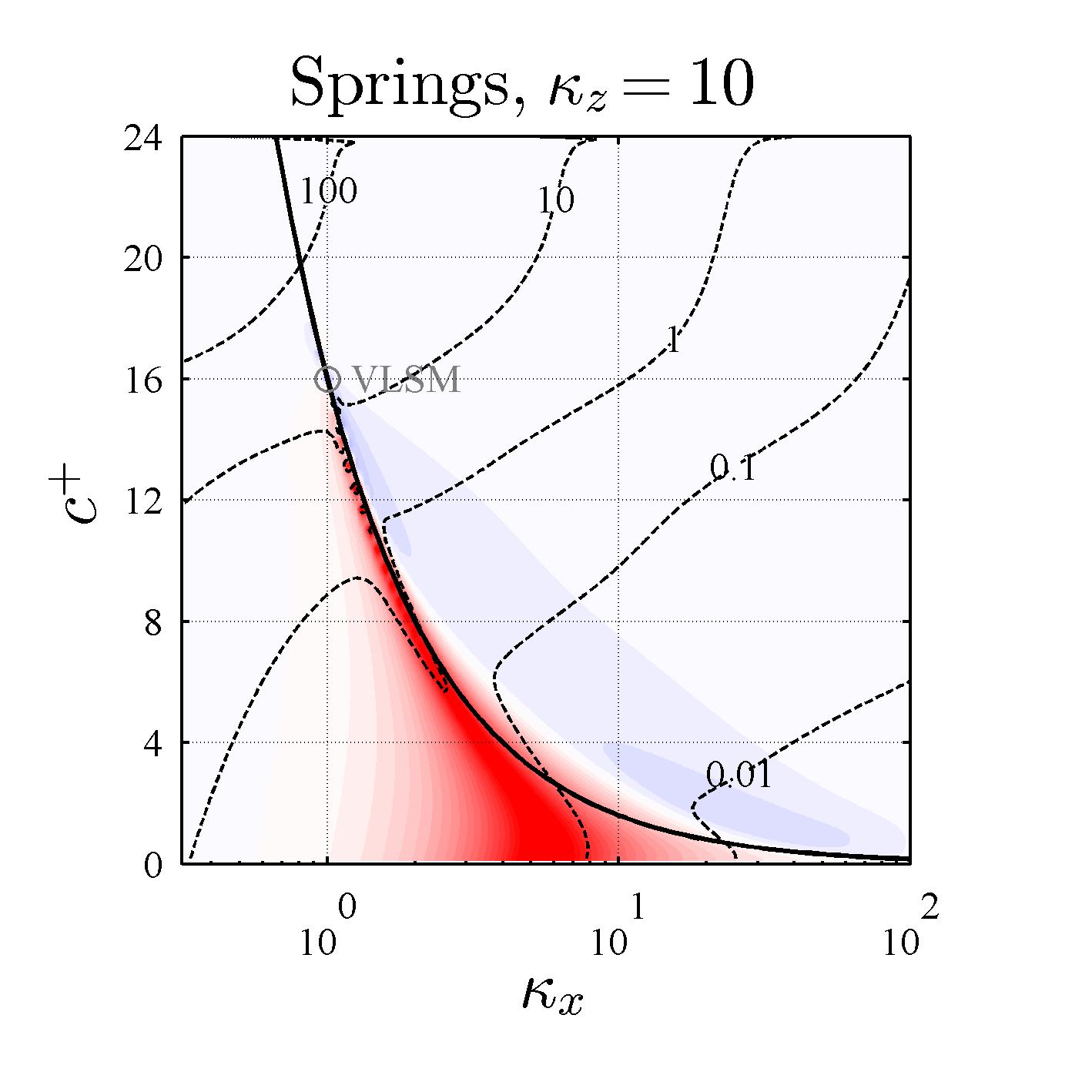}}}%
	\subfigure[]{\resizebox*{4.5cm}{!}{\includegraphics{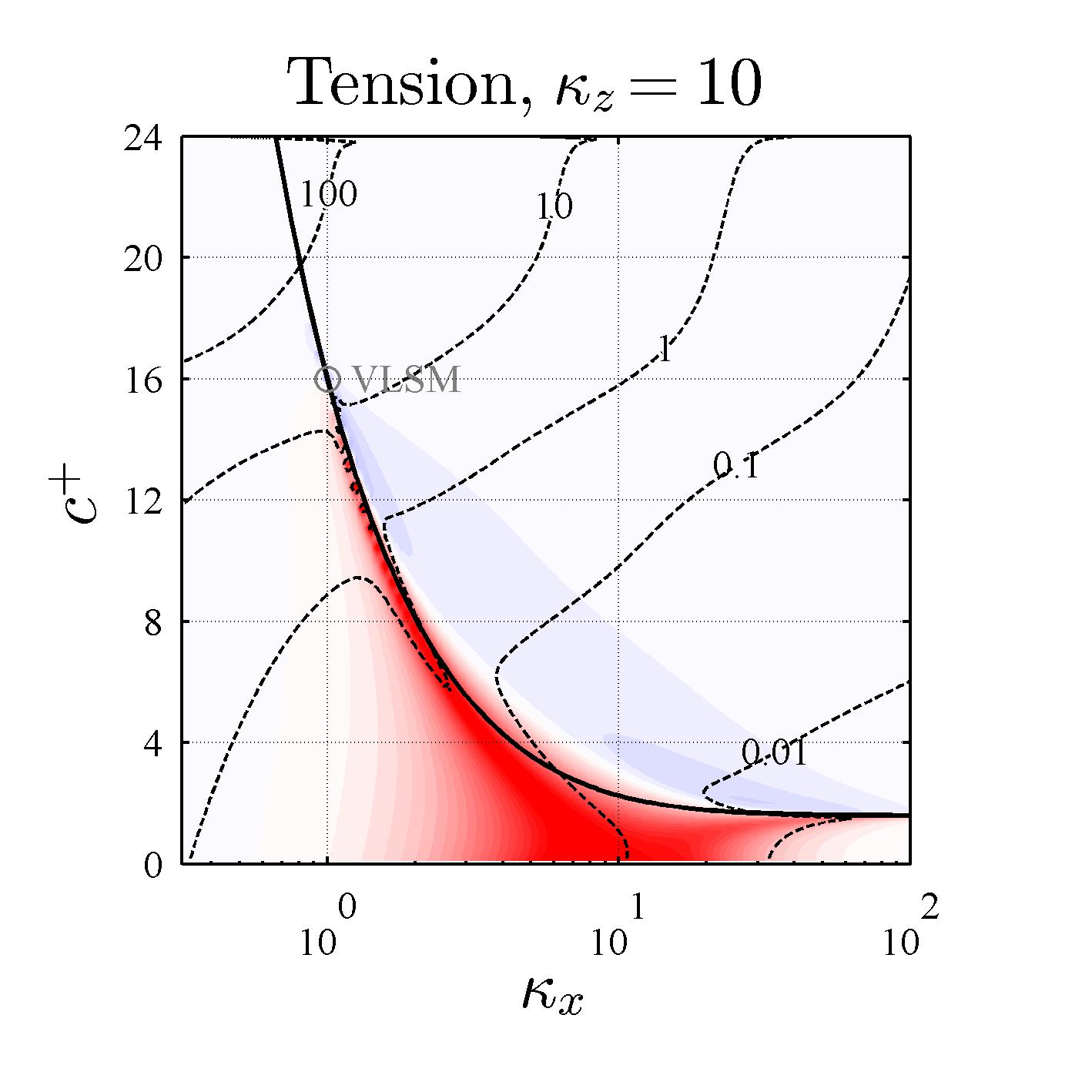}}}%
	\subfigure[]{\resizebox*{4.5cm}{!}{\includegraphics{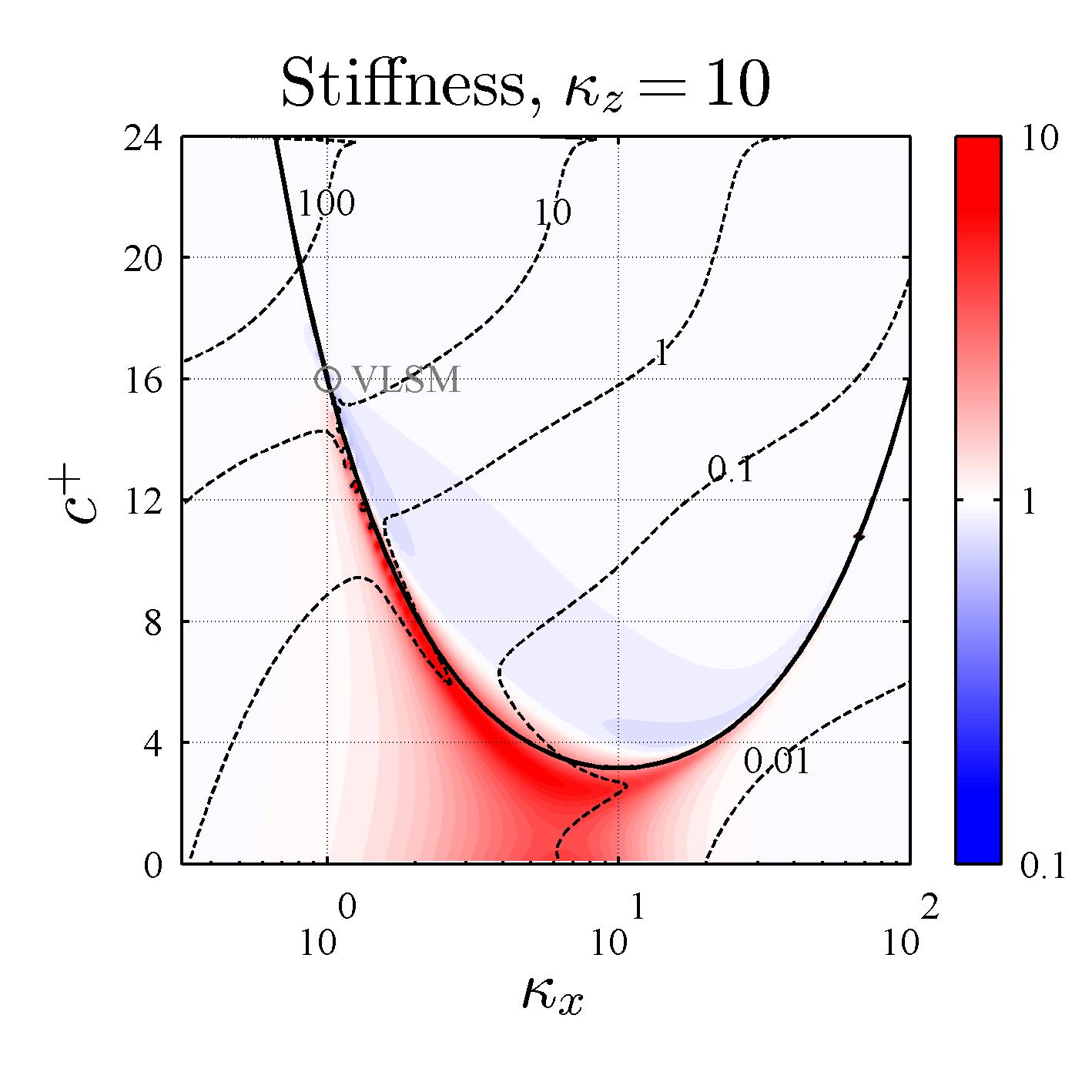}}}\\%
	\subfigure[]{\resizebox*{4.5cm}{!}{\includegraphics{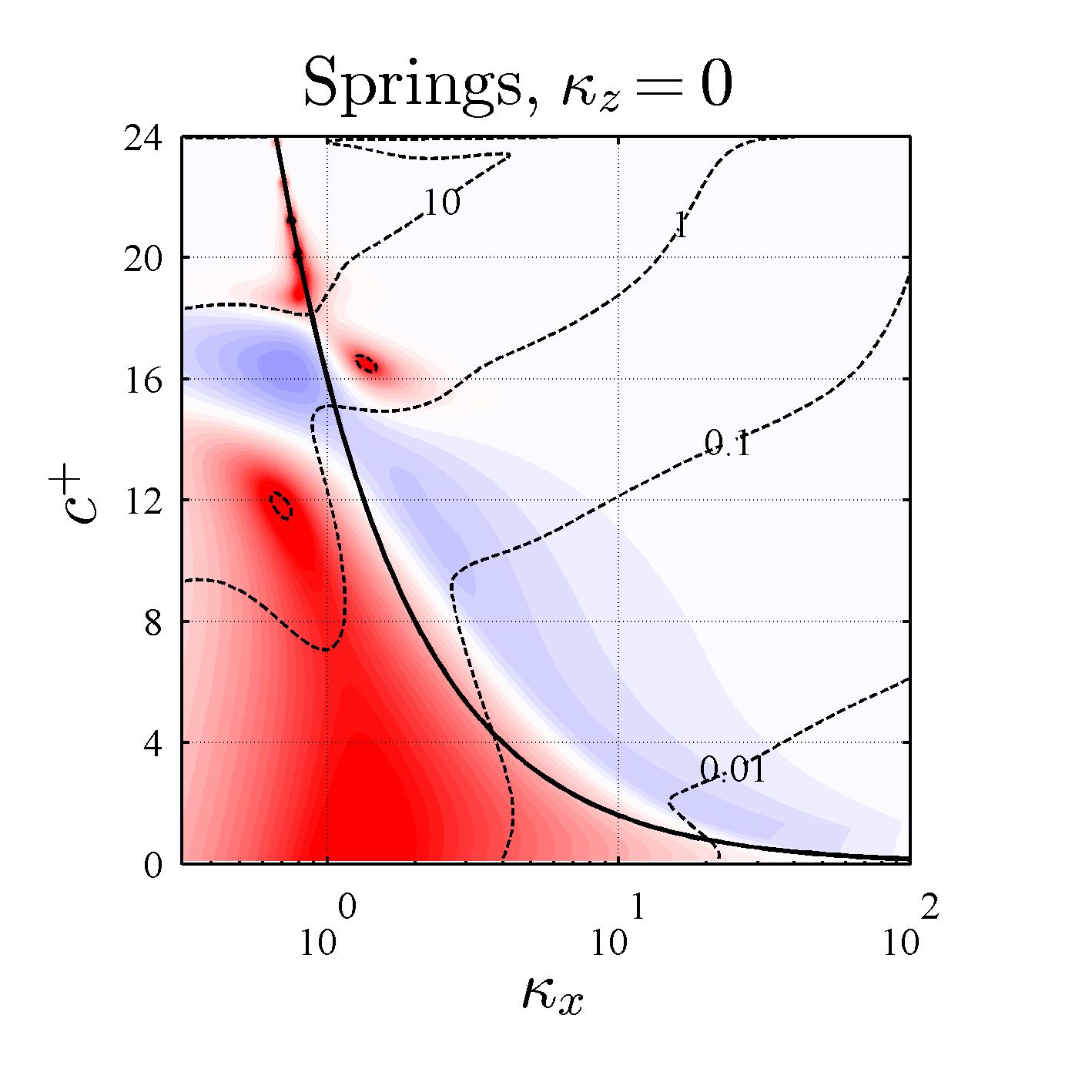}}}%
	\subfigure[]{\resizebox*{4.5cm}{!}{\includegraphics{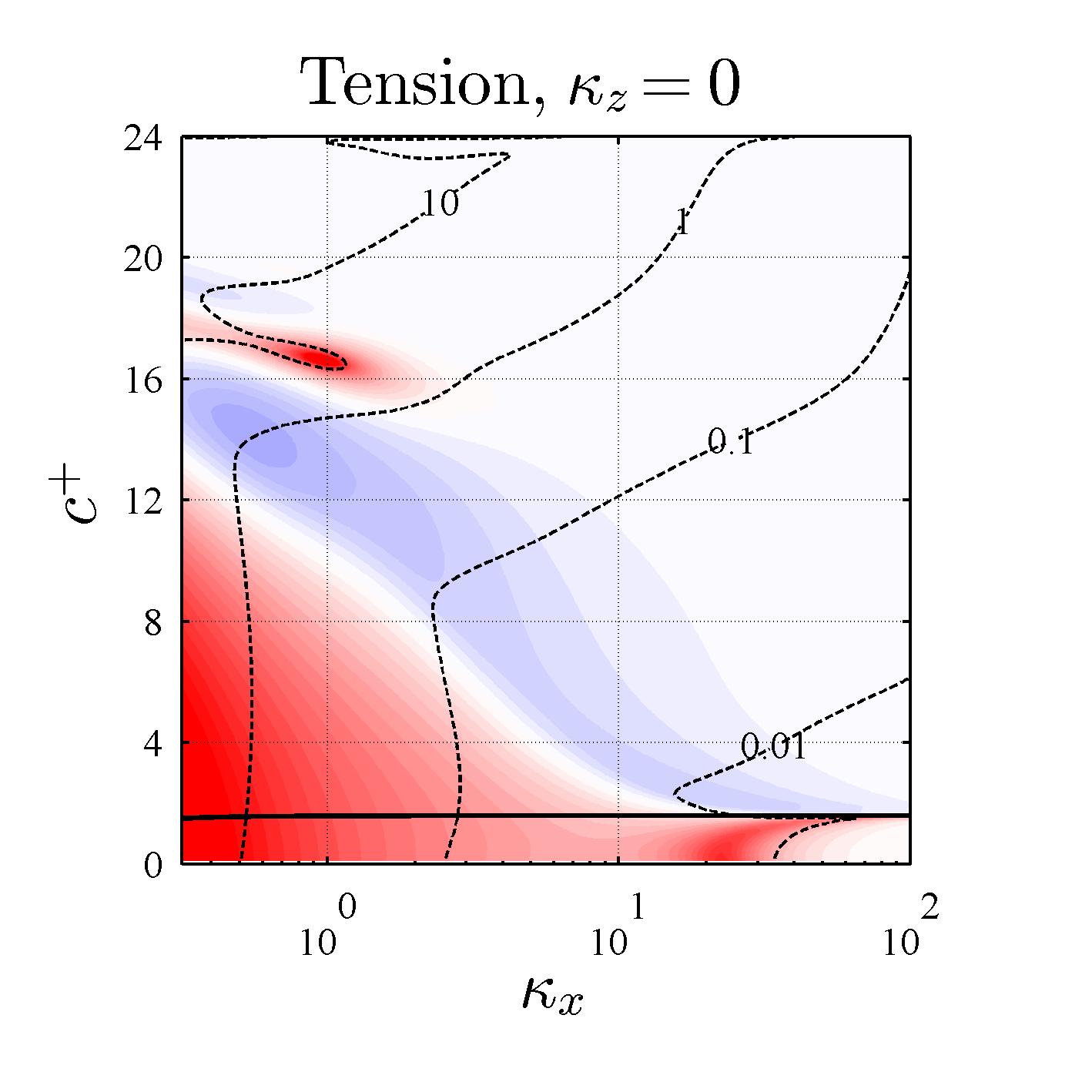}}}%
	\subfigure[]{\resizebox*{4.5cm}{!}{\includegraphics{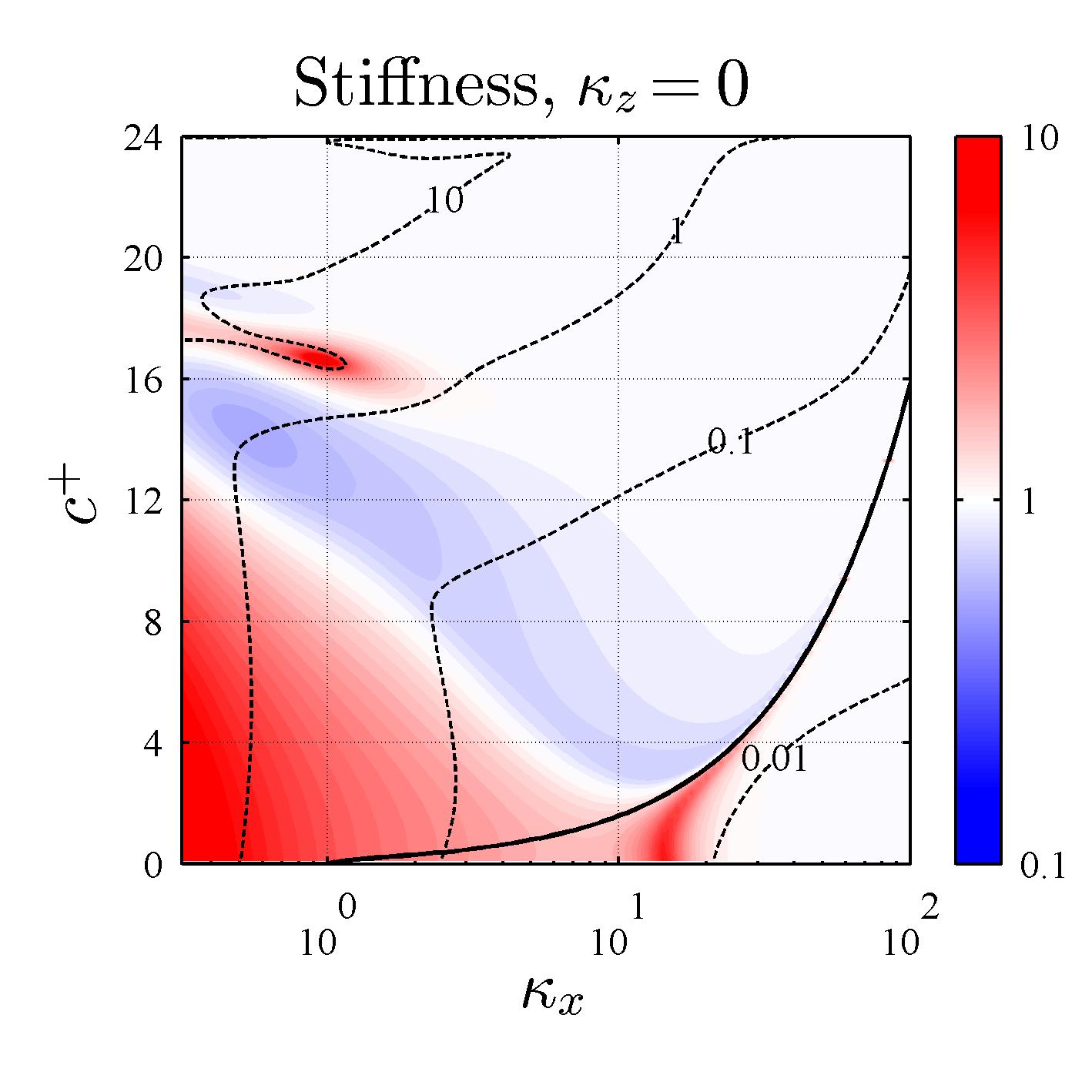}}}%
	\caption{Shaded contours showing the singular value ratio $\skc/\sko$ as a function of streamwise wavenumber and mode speed.  Blue regions denote mode suppression and red regions denote amplification.  Plots (a,d), (b,e) and (d,f) represent the base case, tensioned wall and stiff wall listed in Table~\ref{tab:walls}, respectively.  The dashed contours indicate the magnitude of the singular values $\skc$ over the compliant walls.  The solid lines show the resonant frequency.}
	\label{fig:KTS}
\end{center}
\end{figure}

Next we compare the effects of a compliant wall on a spring support with a tensioned membrane and a stiff plate.  For the simple spring-damper system, the fluid-structure interactions are dependent solely on frequency.  Moreover, the wall does not communicate in the streamwise and spanwise directions, which means that it cannot support wave propagation.  In contrast, tensioned membranes and stiff plates have a wavenumber-dependent effective spring constant (Eq. \ref{eqEffectiveSpring}) and can support wave propagation.  This means that the three different walls have varying effects across spectral space, despite being optimized to suppress the VLSM-type modes.

The above effects are best understood in terms of the resonant frequency $\om_r$.  Like the results shown in the previous section, for $\kz = 10$, modes with frequencies below the resonant frequency are further amplified by the compliant walls while modes with higher frequencies are generally suppressed (Fig.~\ref{fig:KTS}a-c).  However, the resonant frequency (solid black lines) varies substantially across the three different cases.  For the basic spring-damper wall, the resonant frequency is constant, and so the effect of the wall is centered around modes with $c^+ \kx = \om_r$, or $c^+ \sim \kx^{-1}$ (Fig.~\ref{fig:KTS}a).  For the tensioned membrane and stiff plate, the response is centered around a similarly decreasing function $c^+ = f(\kx)$ for $\kx \ll \kz(=10)$ (Fig.~\ref{fig:KTS}b,c).  This is because the effective spring constant is dominated by the spanwise wavenumber dependence for $\kx \ll \kz$, leading to essentially constant $C_{ke}\approx C_t \kz^2 \approx 505$ and $C_{ke} \approx C_s \kz^4 \approx 500$ for the results shown in Fig.~\ref{fig:KTS}b,c.  However, as $\kx \gg \kz$, the effective spring constant for the walls is dominated by the streamwise dependencies $C_{ke}\approx C_t \kx^2$ and $C_{ke}\approx C_s \kx^4$, which translates into resonant frequencies $\om_r \approx  \kx \sqrt{C_t/C_m}$ and $\om_r \approx \kx^2 \sqrt{C_s/C_m}$.   This means that the maximum admittance is found at near-constant $c^+ \approx \sqrt{C_t/C_m} = 1.59$ for the tensioned membrane (Fig.~\ref{fig:KTS}b) and is an increasing function $c^+ = f(\kx)$ for the stiff plate (Fig.~\ref{fig:KTS}c).

Figures~\ref{fig:KTS}d-f show that all three walls also lead to significant amplification of certain classes of two-dimensional ($\kz = 0$) structures, which is consistent with previous experiments and DNS.  Interestingly, all three cases exhibit a repeating amplification-suppression pattern across spectral space.  As an example, for fixed phase speed $c^+ \approx 10$, long structures with streamwise wavelength $\kx < 1$ are further amplified, while shorter modes with $\kx > 2$ are suppressed over the compliant walls.  The wavenumber at which this amplification-suppression transition occurs generally decreases with increasing $c^+$ and there is an additional suppression-amplification transition at higher speeds (see \eg $c^+ \approx 16$ in Fig.~\ref{fig:KTS}d-f), although wall resonance also plays an important role (Fig.~\ref{fig:KTS}d-f, solid black lines).  In general, there appear to be two classes of mode that are further amplified over compliant walls.  Long, slow-moving modes with $\kx < 5$ and $c^+ < 7$ are amplified regardless of the wall properties, at least for the walls tested.  The second class of modes that is further amplified is linked to wall resonance and is generally of smaller wavelength (see \eg $\kx > 10$ in Fig.~\ref{fig:KTS}e,f).

Note once again that the resonant frequency, and hence wave speed, varies significantly across the three different cases.  The wave speed corresponding to resonance is a decreasing function of $\kx$ for the spring-damper wall (Fig.~\ref{fig:KTS}d), constant for the tensioned membrane (Fig.~\ref{fig:KTS}e, $c^+ \approx \sqrt{C_t/C_m}$, \ie the free-wave speed of the wall), and an increasing function of $\kx$ for the stiff plate (Fig.~\ref{fig:KTS}f).

\subsection{Anisotropy and Wall-based Instability}\label{sec:results-anisotropy}

\begin{figure}
\begin{center}
	\subfigure[]{\resizebox*{5cm}{!}{\includegraphics{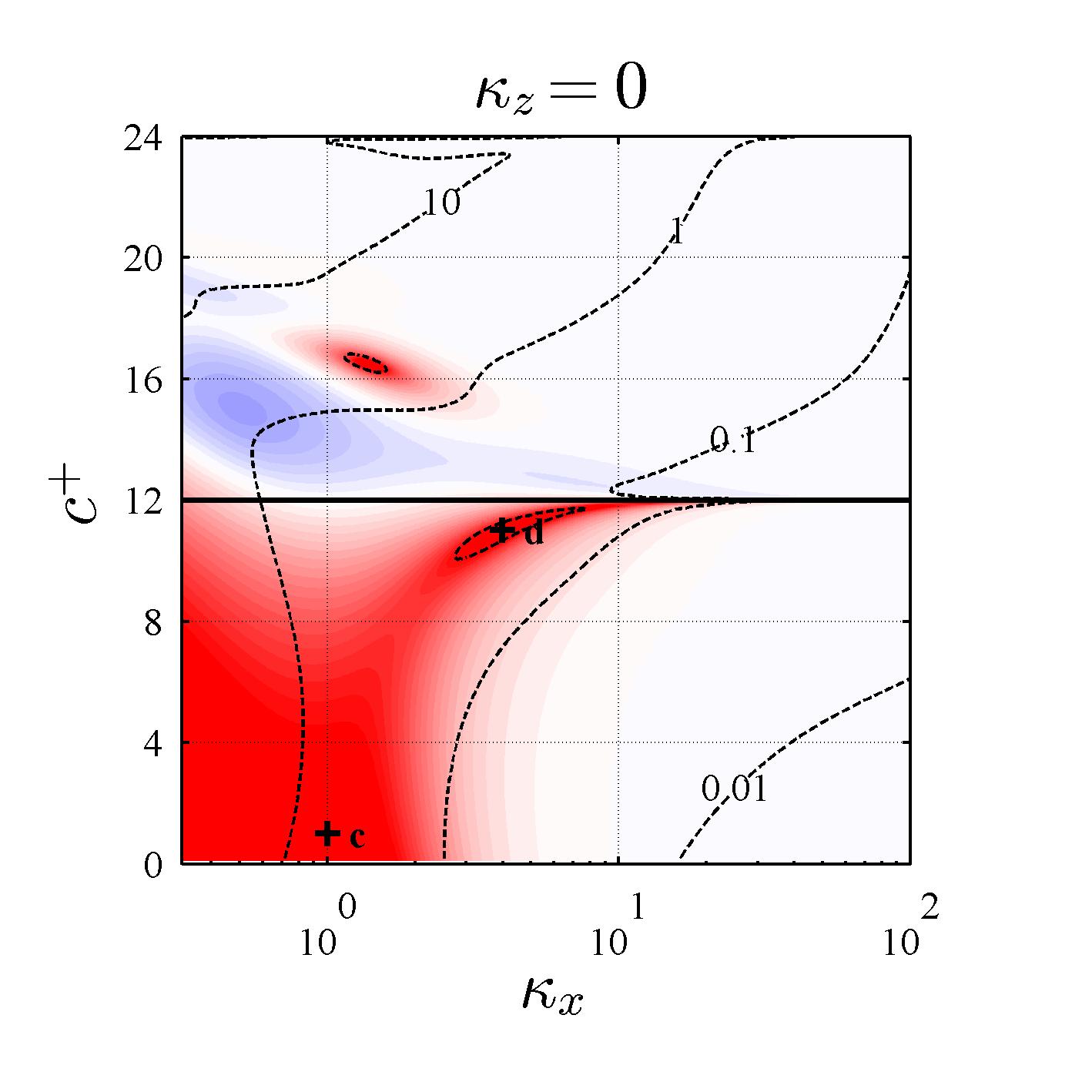}}}%
	\subfigure[]{\resizebox*{5cm}{!}{\includegraphics{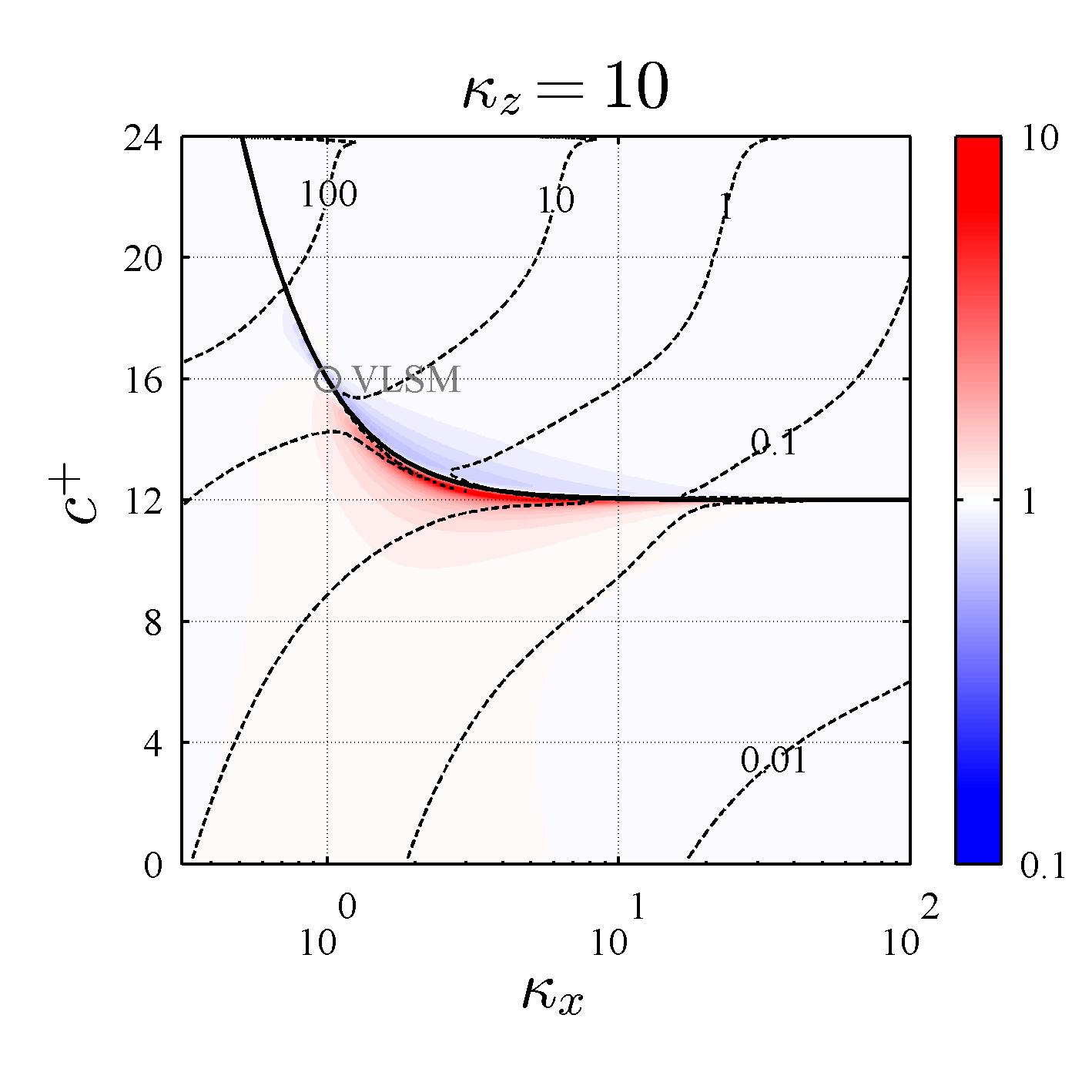}}}\\%
	\subfigure[]{\resizebox*{10cm}{!}{\includegraphics{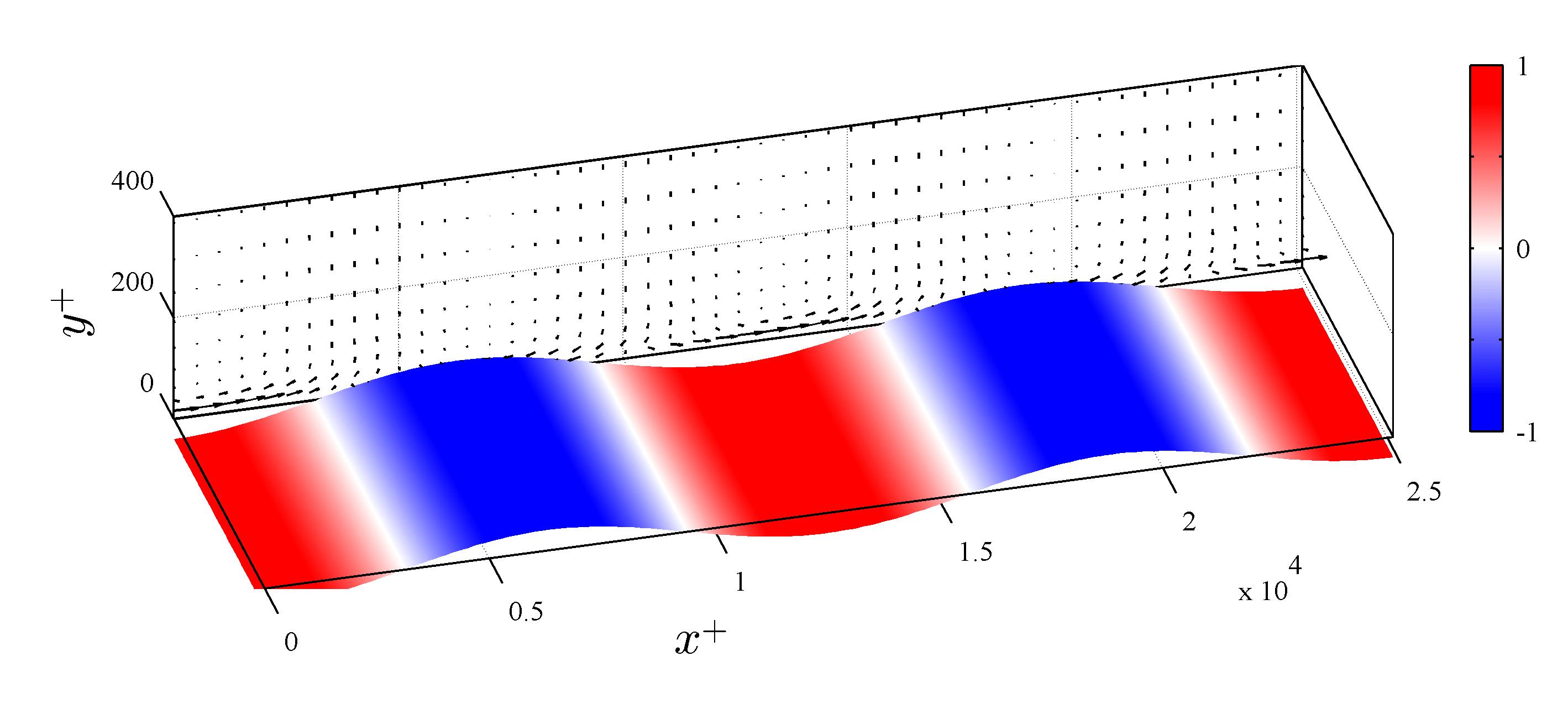}}}\\%
	\subfigure[]{\resizebox*{10cm}{!}{\includegraphics{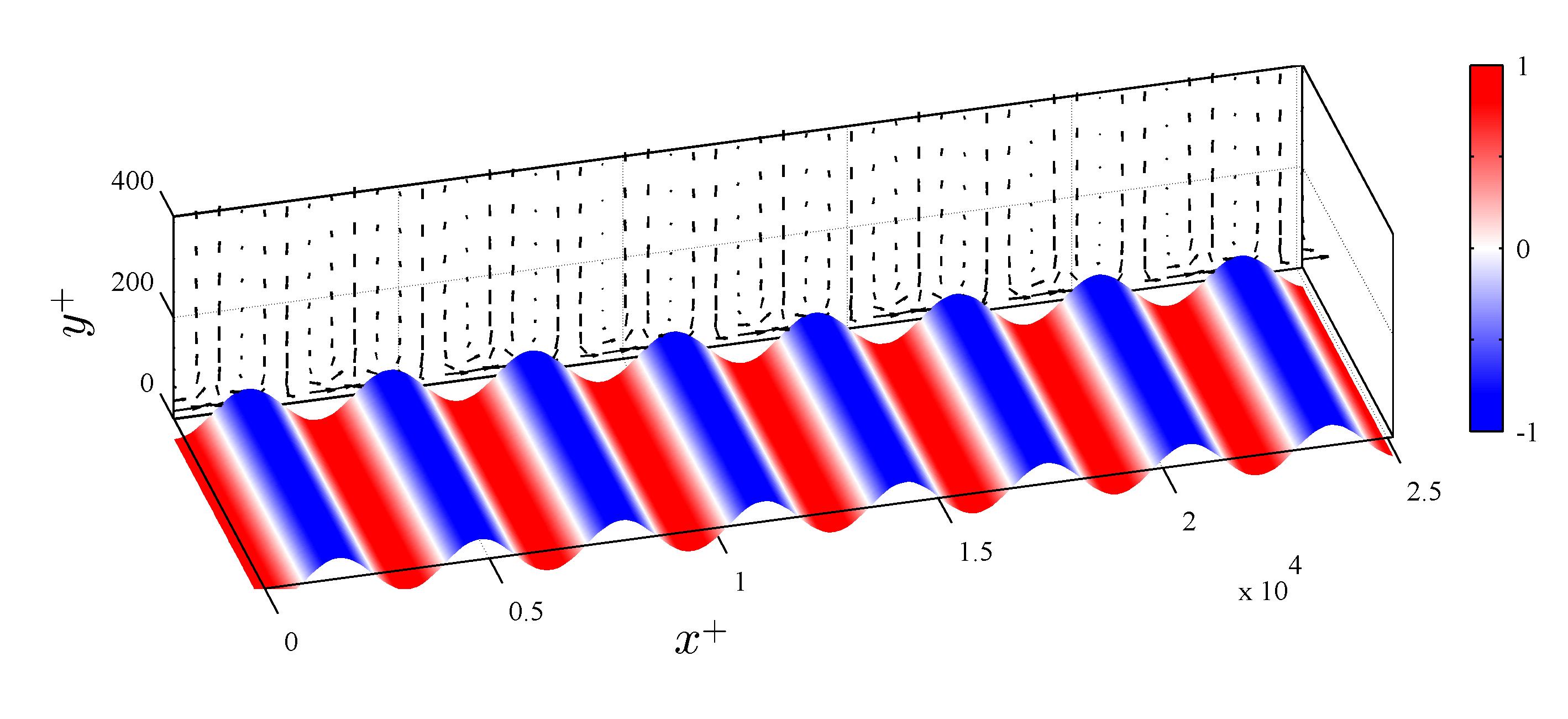}}}%
	\caption{Shaded contours showing the singular value ratio $\skc/\sko$ for $\kz = 0$ (a) and $\kz = 10$ (b) over the anisotropic wall listed in Table~\ref{tab:walls}.  Blue regions denote mode suppression and red regions denote amplification.  Plots (c) and (d) show the structure of the highly amplified two-dimensional modes marked in (a), representing wave number frequency combinations $\kb = (\kx,\kz,c^+) = (1,0,1)$ and $\kb = (3.8,0,11)$, respectively.  The shading on the compliant wall indicates the normalized pressure field.  The vectors show the streamwise and wall-normal velocity fields.  Wall deflection not to scale.}
	\label{fig:Structure}
\end{center}
\end{figure}

In this section, we introduce the effects of anisotropy by testing the effects of a wall with different streamwise and spanwise tension coefficients $C_{tx} = 288$ and $C_{tz} = 2.224$, so that $C_{ke} = C_{tx}\kx^2 + C_{tz}\kz^2$ (Eq.~\ref{eqEffectiveSpring}).  This anisotropy changes the resonant frequency of the wall (Fig.~\ref{fig:Structure}a,b) and the free wave speed is now $c^+ \approx \sqrt{C_{tx}/C_m} = 12$.  However, the trends observed in the previous section remain.  In particular, there is a sharp transition in performance across the resonant frequency for the $\kz = 10$ modes, and spanwise-constant ($\kz=0$) modes are susceptible to significant further amplification.  There are two classes of highly-amplified spanwise-constant modes: long, slow-moving structures (\eg $\kx =1, c^+ = 1$, marked $c$ in Fig.~\ref{fig:Structure}a) and shorter, faster structures moving at close to the free wave speed (\eg $\kx \approx 4, c^+ = 11$, marked $d$ in Fig.~\ref{fig:Structure}a).

The above predictions are broadly consistent with the observations of Gad-el-Hak et al., \cite{GadelHak1984,GadelHak1986}, who showed that elastic and viscoelastic layers under turbulent boundary layers gave rise to two distinct classes of surfaces waves: the first, termed static divergence, were very long, slow-moving (nearly static) structures, while the second class of surface waves had shorter wavelengths and faster phase speeds, comparable to the free shear wave speed of the layer.  The experiments suggest that the static-divergence waves appear preferentially for viscoelastic coatings while the faster waves appear preferentially for elastic layers.  This effect of the viscosity (\ie the damping in our model) remains to be explored.

Figures~\ref{fig:Structure}c,d show the structure associated with the two highly-amplified modes identified in Fig.~\ref{fig:Structure}a.  Although the resolvent modes have vastly different wavelengths and speeds, the overall structure is similar.  Specifically, the streamwise velocities associated with the modes are confined to a very small layer close to the wall, above which the velocities are primarily in the up-down wall-normal direction.  Further, the magnitude of the wall-pressure field is largest over surface troughs and smallest over surface peaks, \ie high pressures coincide with downward deflections and vice versa, as expected physically.

\subsection{Reynolds Number Effects}\label{sec:results-Reynolds}
\begin{figure}
	\begin{center}
		\subfigure[]{\resizebox*{5cm}{!}{\includegraphics{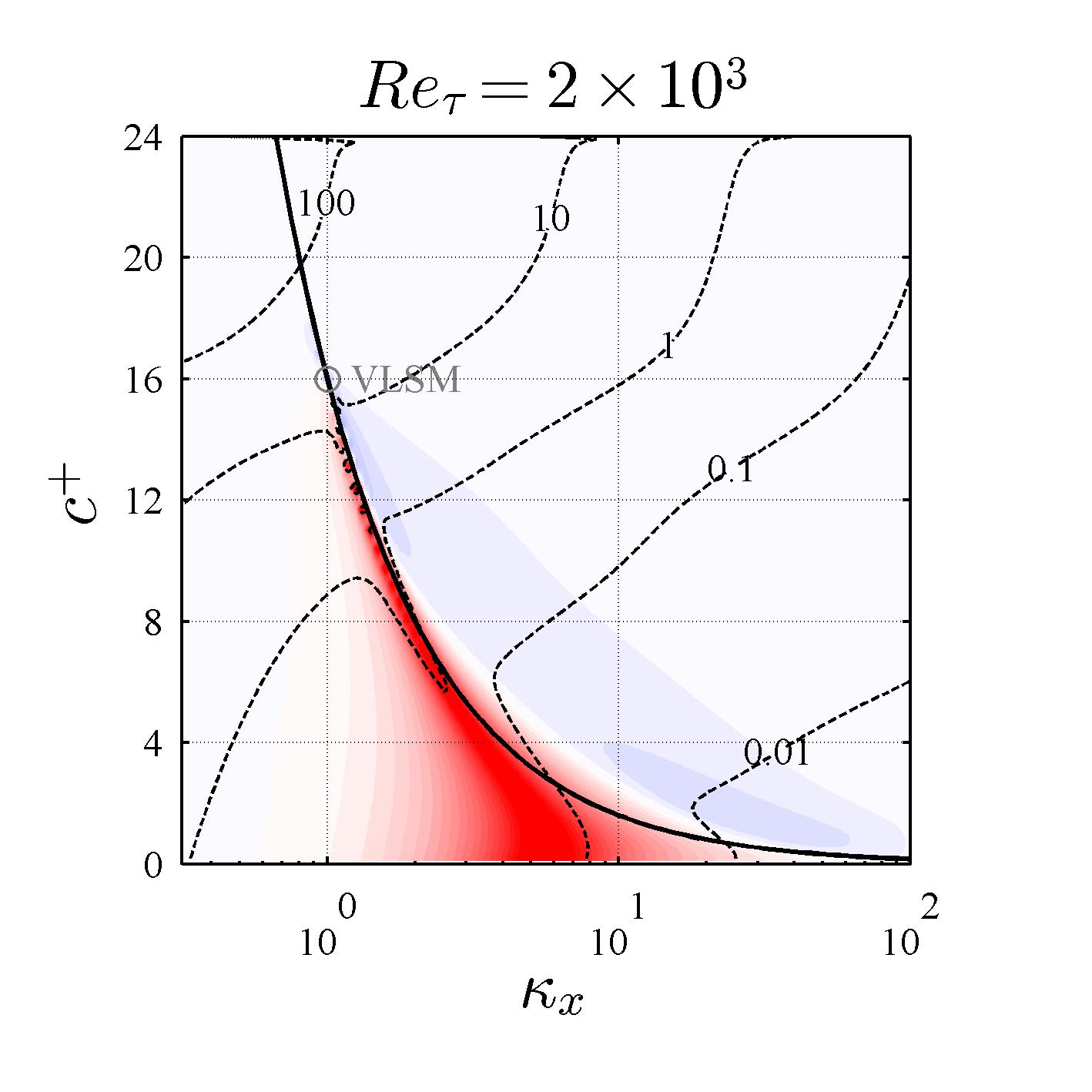}}}%
		\subfigure[]{\resizebox*{5cm}{!}{\includegraphics{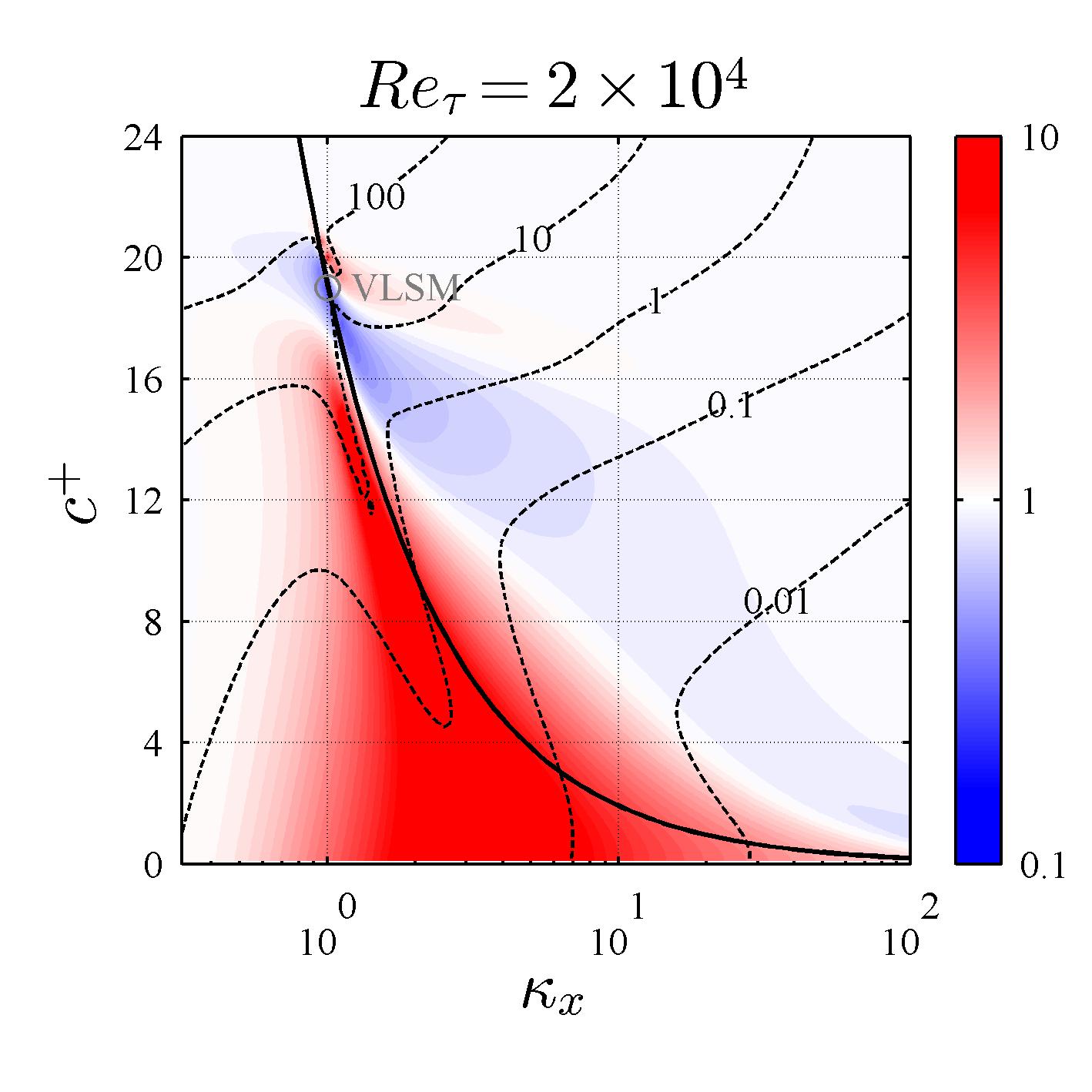}}}%
		\caption{Shaded contours showing the ratio of compliant wall to null-case singular values, $\skc/\sko$, at friction Reynolds number $Re_\tau = 2\times 10^3$ (a) and $Re_\tau = 2\times 10^4$ (b) for spring-damper walls optimized to suppress VLSM-type structures.  Blue regions denote mode suppression while red regions indicate further amplification.  The solid black lines indicate the resonant frequency.  All results correspond to $\kz = 10$.}
		\label{fig:Re}
	\end{center}
\end{figure}

In this section, we consider Reynolds number effects.  Specifically, we contrast the effect of walls optimized to suppress VLSM-type modes at $\Ret = 2000$ and $\Ret = 2 \times 10^4$, characterized by wavenumber-frequency combinations $\kb = (\kx,\kz,c^+)=(1,10,16)$ and $\kb = (1,10,19)$, respectively.  The assumed increase in mode speed with Reynolds number is consistent with the $y^+ \sim \sqrt{\Ret}$ scaling for such large-scale modes proposed previously \cite{Marusic2010}.

Similar to the low mass ratio case (Fig.~\ref{fig:Cm}a), Fig.~\ref{fig:Re}b shows that the compliant wall optimized for $\Ret = 2\times 10^4$ has a larger region of influence in spectral space.  Physically, this is because, at higher $\Ret$, the optimization targets a faster moving mode with $c^+ = 19$ that is centered further away from the wall.  Since slower-moving modes with similar wavenumbers (\ie similar length scales) are likely to be centered closer to the wall and therefore have higher wall pressure signatures, the compliant wall also interacts strongly with them.

Note that the amplification-suppression patterns observed previously (Fig.~\ref{fig:Cm}-Fig.~\ref{fig:KTS}) also persist at higher $\Ret$.  For example, at $c^+ = 12$, longer modes with $\kx <2$ are further amplified over the compliant wall while shorter modes with $\kx >3$ are suppressed.  For higher mode speeds $c^+ > 18$, this pattern reverses whereby longer modes with $\kx < 1$ are suppressed and shorter modes are amplified.  A comparison of Fig.~\ref{fig:Re}a and \ref{fig:Re}b indicates that the general patterns of mode suppression and amplification remain broadly similar, with Reynolds number and wall resonance serving to shift the transition points.  Importantly, this observation suggests that it may be possible to generate scaling laws for compliant wall performance that are useful for all Reynolds numbers.  Previous work shows that the structure and amplification of smooth-wall resolvent modes exhibit distinct Reynolds-number scaling regimes depending on whether the modes are centered in the near-wall, logarithmic, or outer region of the flow \cite{Moarref2013}.  As such, it is perhaps not surprising that the effects of passive or active control also appear to scale predictably with Reynolds number.

\subsection{Optimal Wall from Fukagata et al. 2008}\label{sec:results-F2008}

\begin{figure}
	\begin{center}
		\subfigure[]{\resizebox*{5cm}{!}{\includegraphics{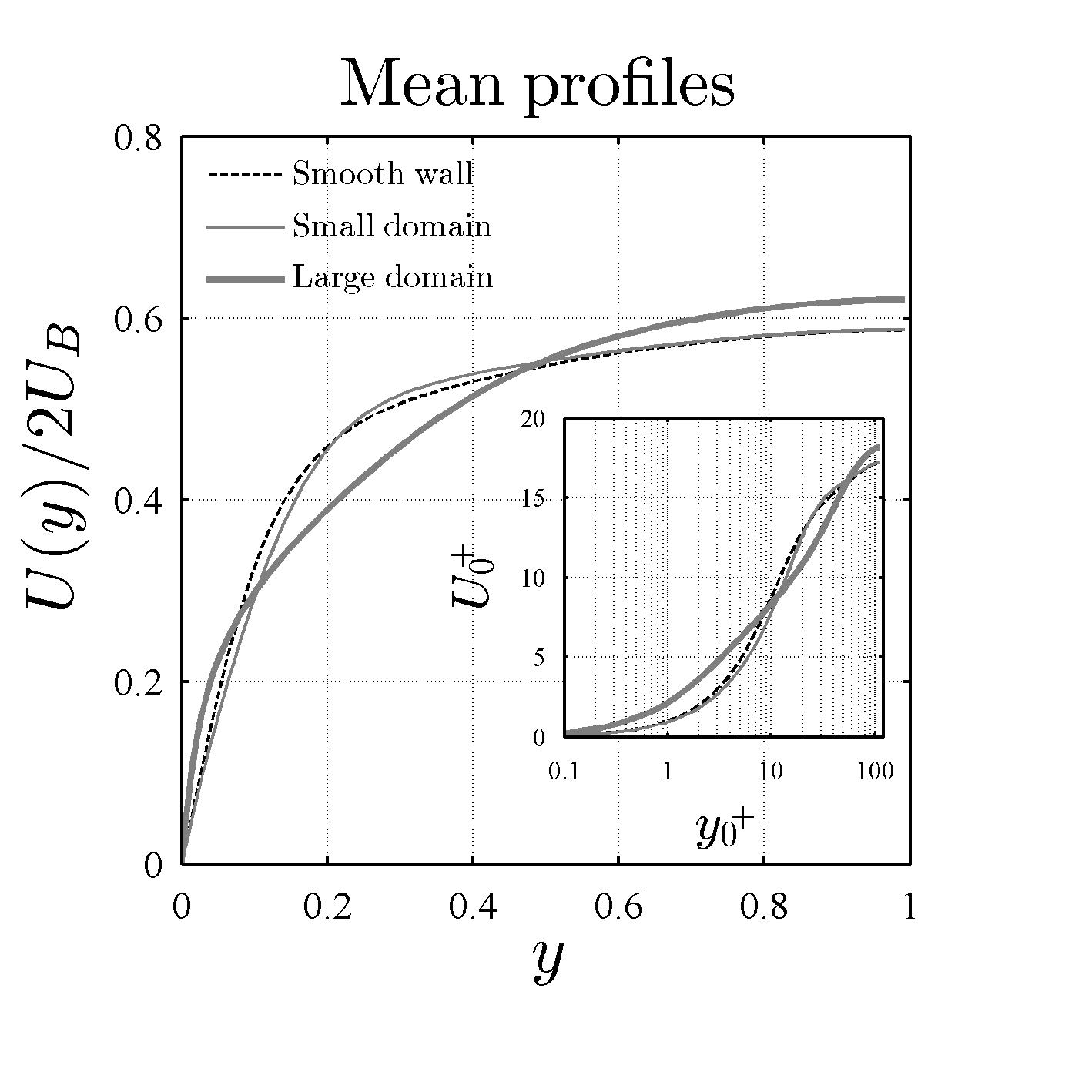}}}%
		\subfigure[]{\resizebox*{5cm}{!}{\includegraphics{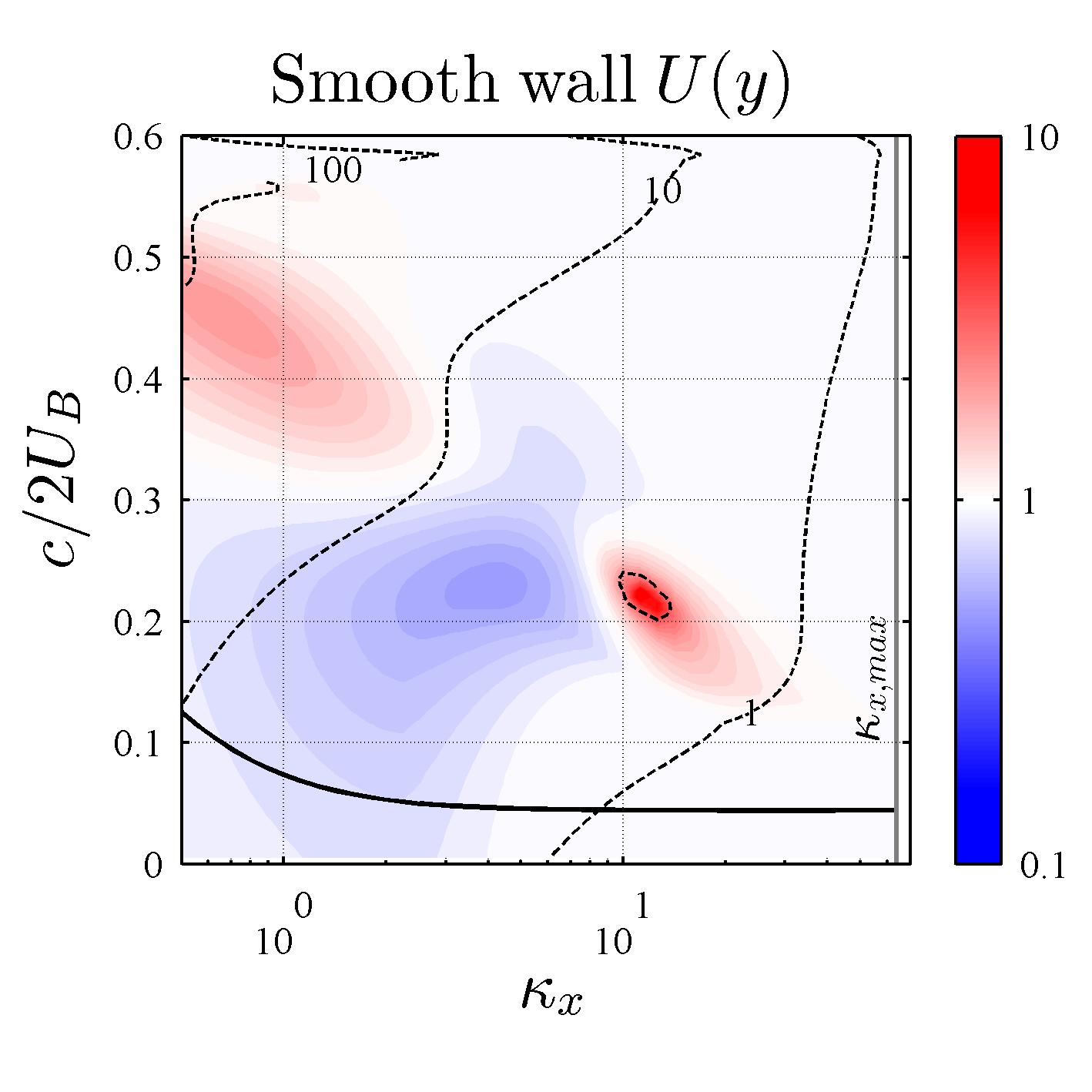}}}\\%
		\subfigure[]{\resizebox*{5cm}{!}{\includegraphics{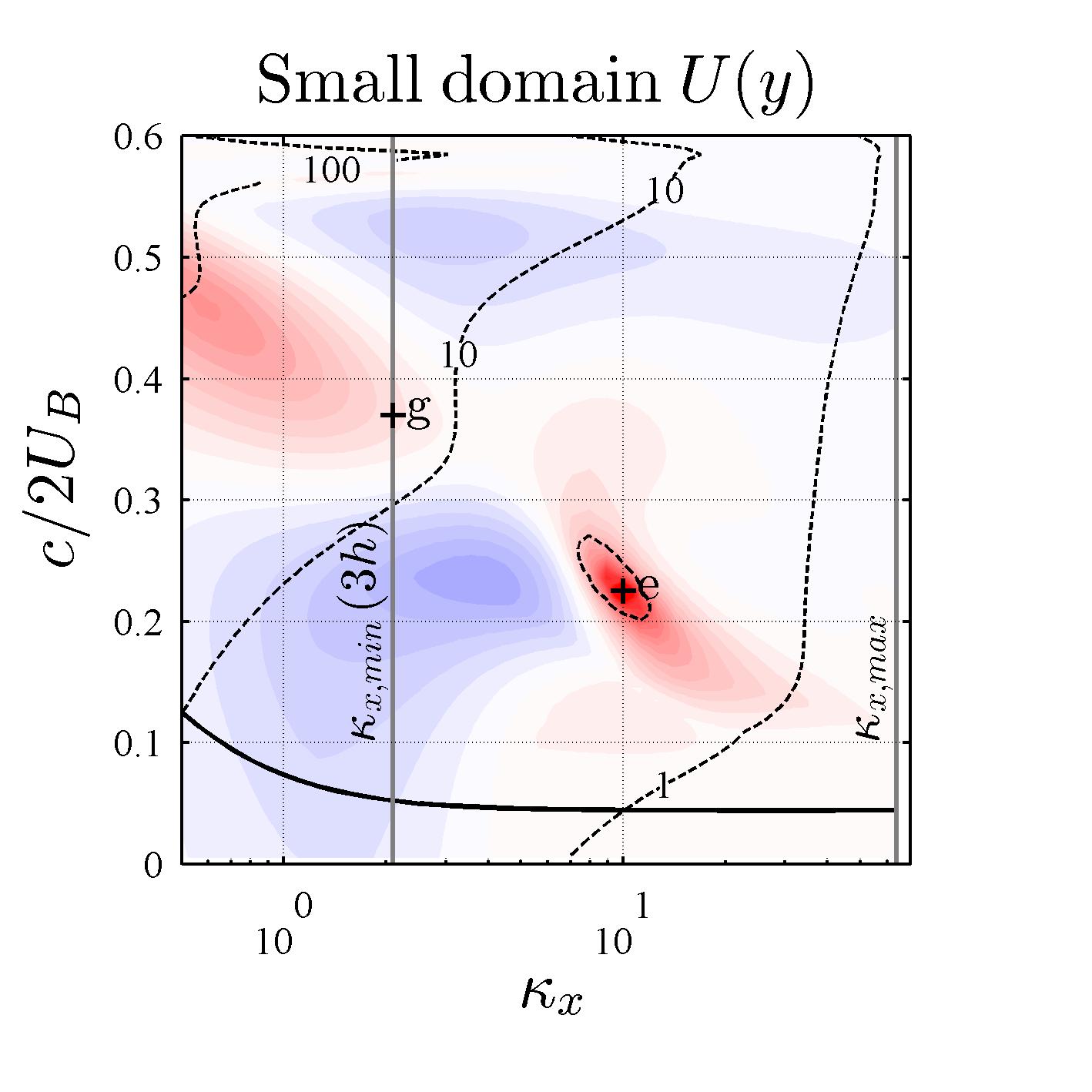}}}%
		\subfigure[]{\resizebox*{5cm}{!}{\includegraphics{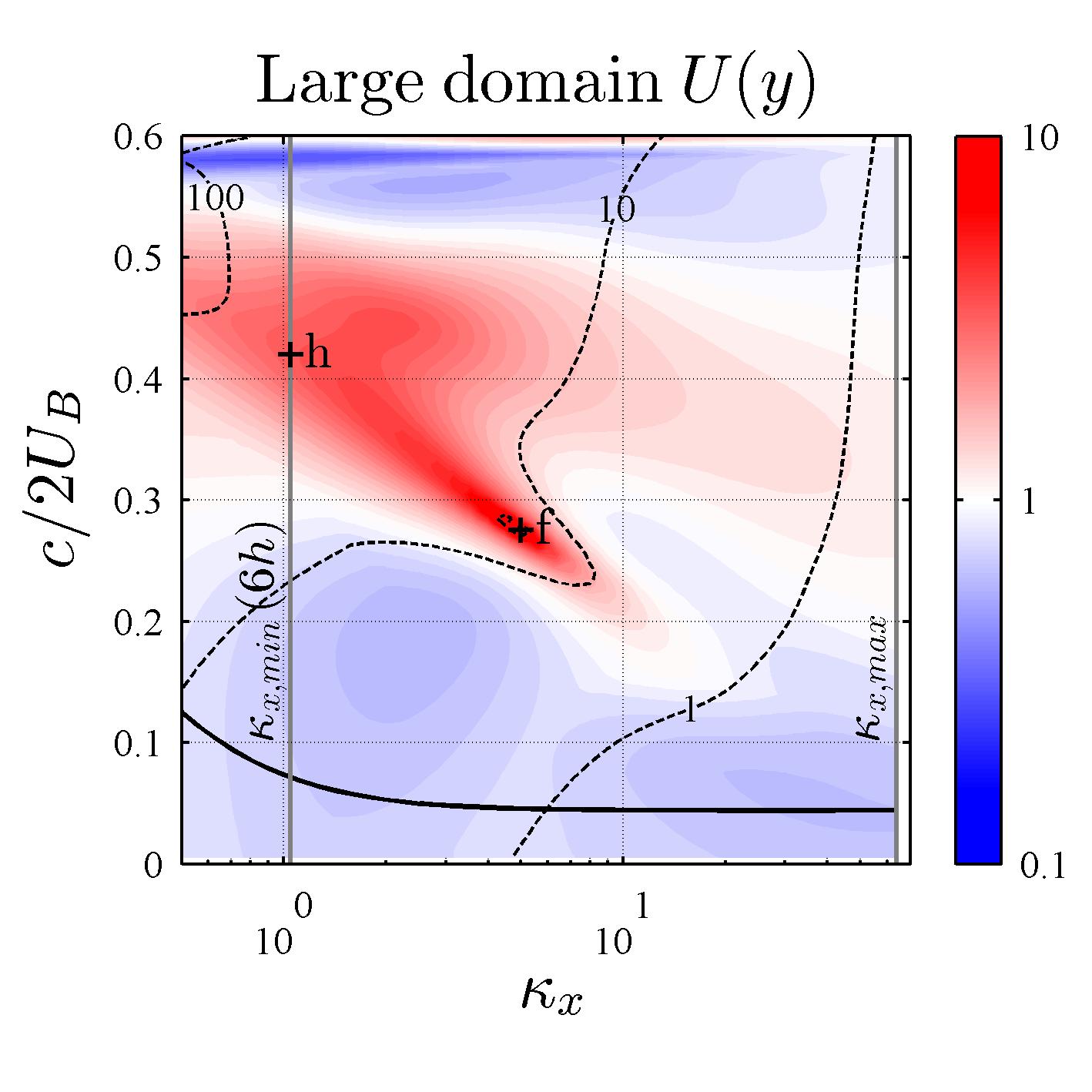}}}\\%
		\subfigure[]{\resizebox*{6.5cm}{!}{\includegraphics{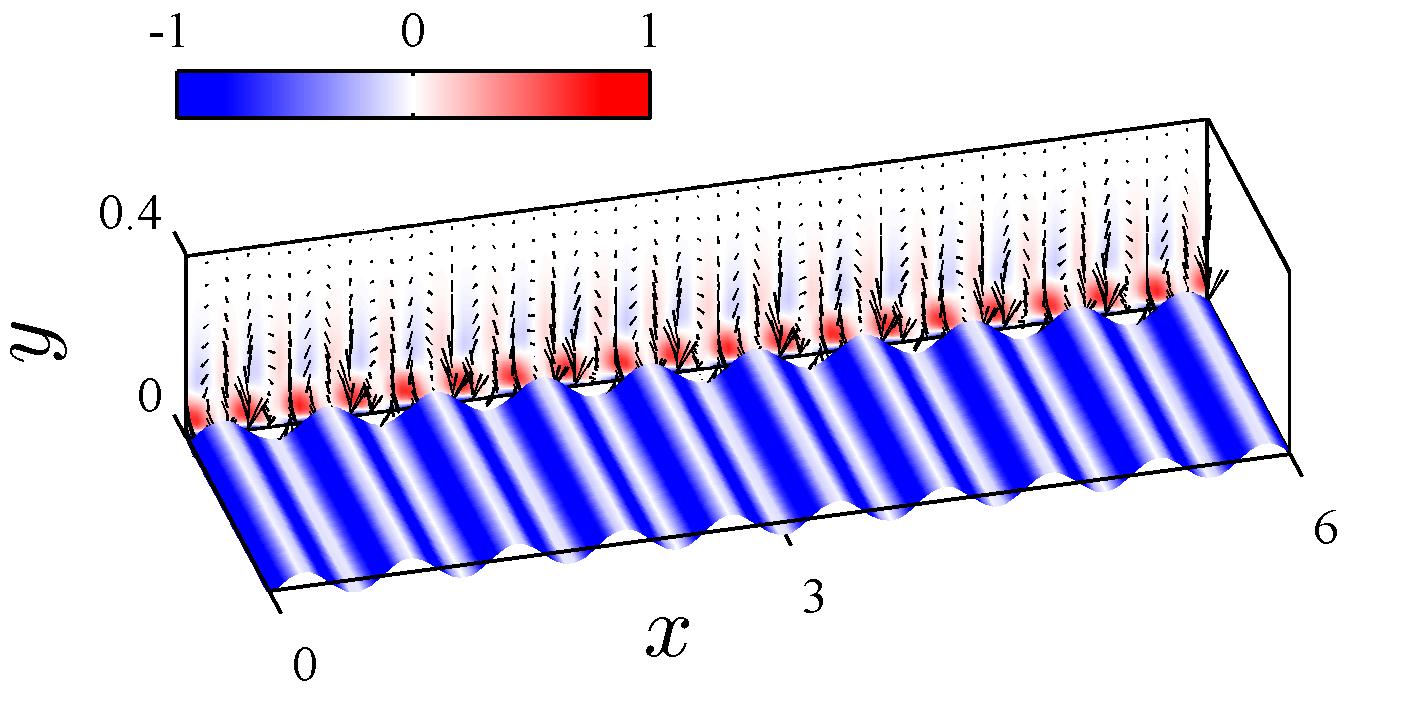}}}%
		\subfigure[]{\resizebox*{6.5cm}{!}{\includegraphics{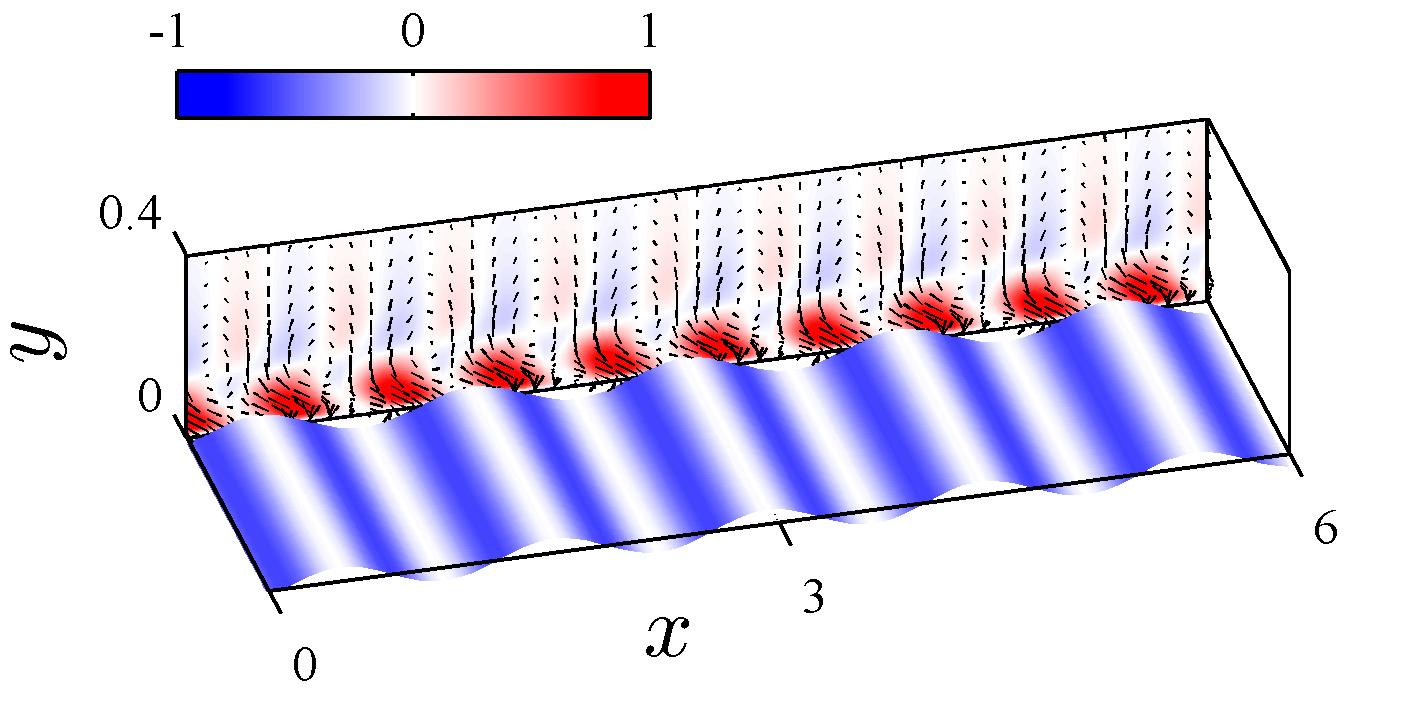}}}\\%
		\subfigure[]{\resizebox*{6.5cm}{!}{\includegraphics{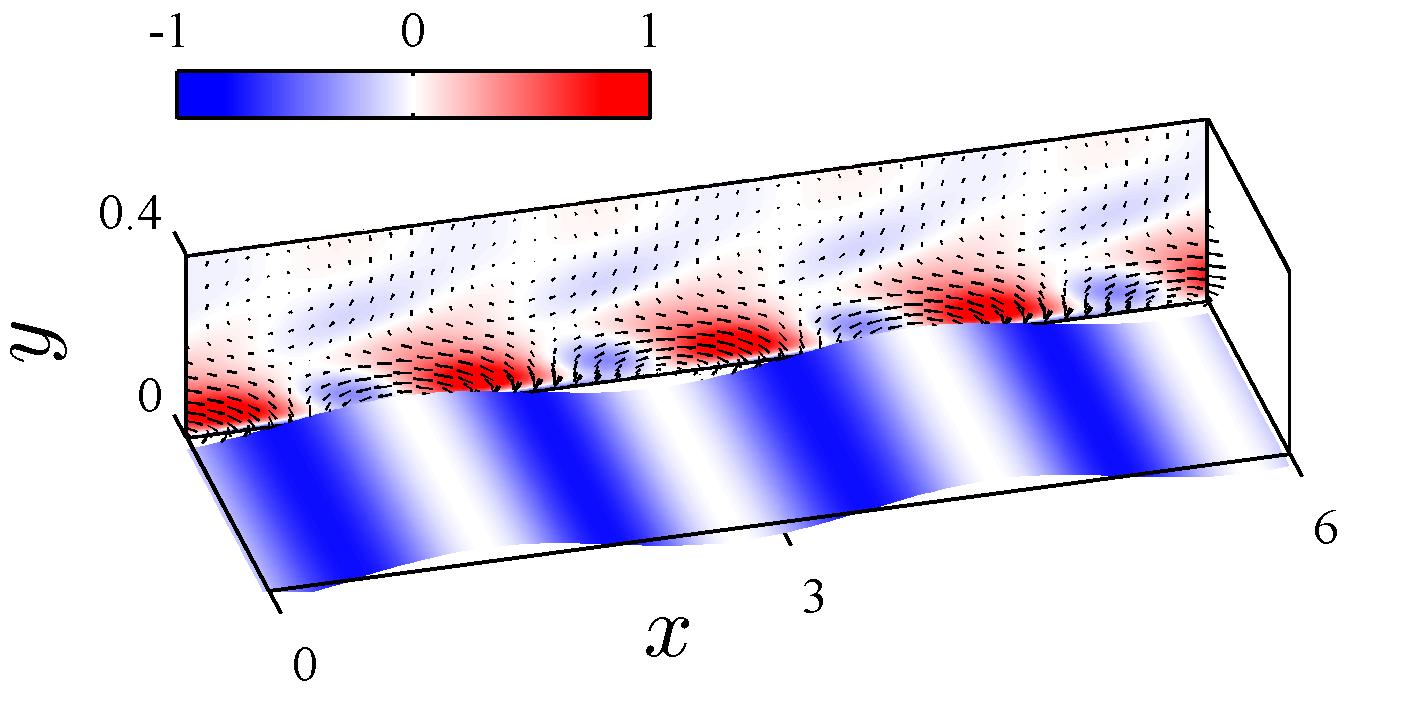}}}%
		\subfigure[]{\resizebox*{6.5cm}{!}{\includegraphics{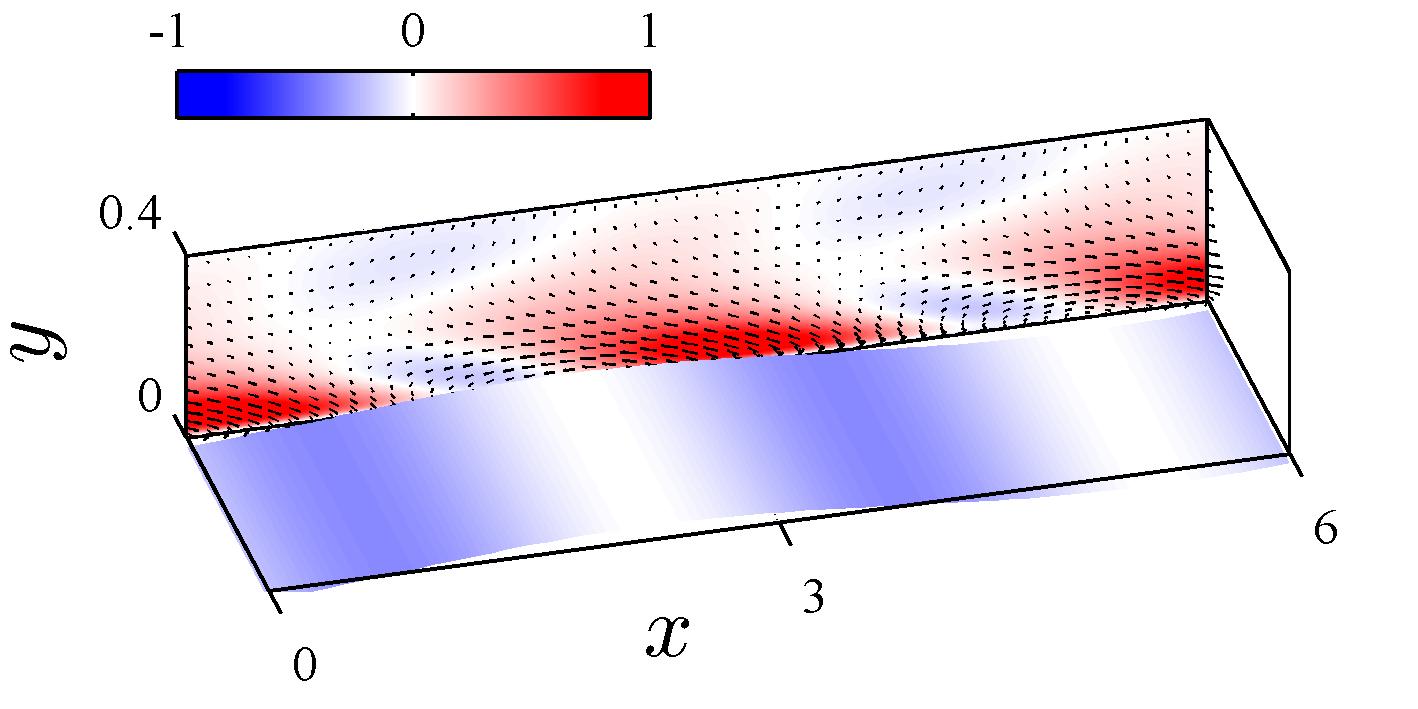}}}%
		\caption{(a) The mean velocity profiles obtained in F2008 \cite{Fukagata2008} over a smooth wall, and compliant walls in small and large domains.  The inset shows the mean profiles plotted in wall units, with the smooth wall friction velocity $u_{\tau 0}$ used for normalization. (b-d) Shaded contours showing the predicted singular value ratio $\skc/\sko$ for spanwise constants modes ($\kz = 0$) over the optimal (A1) compliant wall, but with varying mean velocity profiles used in the resolvent operator.  Blue regions denote mode suppression and red regions denote amplification.  The dashed black lines indicate the actual singular values $\skc$, the solid black line shows the natural frequency of the wall, and the vertical gray lines in (c-d) reflect the minimum cell size and maximum domain length.  Plots (e-g) show the structure of the highly amplified two-dimensional modes marked in (c) and (d).  The shading on the walls shows the normalized fluctuating Reynolds stress field.  The vectors show the streamwise and wall-normal velocity fields.  Wall deflection not to scale.}
		\label{fig:Fukagatakz0}
	\end{center}
\end{figure}

Finally, we present resolvent-based predictions for the optimal anisotropic compliant wall from F2008, focusing primarily on the emergence of the two-dimensional ($\kz=0$) wavelike motions observed in the simulations.  The wall properties listed in Table~\ref{tab:F2008} lead to an effective spring constant of:

\vspace{-0.5cm}
\begin{multline}
C_{ke} = C_k + E b k^2 \sin^2 \theta + \frac{Eb^3}{12(1-\nu_p^2)} k^4 \cos^2 \theta \\ = 4.25 \times 10^{-5} + 2.39 \times 10^{-5} k^2 + 7.08 \times 10^{-11} k^4.
\end{multline}

{\noindent}Since $b = 0.01$, the third term $\propto Eb^3$ is much smaller than the second term $\propto Eb$ for wavenumbers $k \le O(100)$.  Further, the first term is smaller than the second term for $k \ge O(1)$.  Thus, for $1 \ll k \ll 100$, the effective spring constant can be approximated as $C_{ke} \approx 2.39 \times 10^{-5} k^2$, leading to $\om_n = \sqrt{C_{ke}/(\rho_m b)} \approx 0.044 k$.  As illustrated by the solid black lines in Fig.~\ref{fig:Fukagatakz0}, this effective wall stiffness results in a near-constant resonant wave speed of $c/2U_B = \om_n/\kx \approx 0.044$ ($c^+ \approx 1.3$) for modes with $\kz = 0$ and $\kx \approx 3$ to $\kx = 64$.  Further, the \textit{soft} nature (\ie low resonant wave speed) of the wall means that it has a marked influence on singular values across most of the spectral space shown in Fig.~\ref{fig:Fukagatakz0}b-d.

Note that panels b-d in Fig.~\ref{fig:Fukagatakz0} show resolvent-based predictions for mode amplification over the compliant wall made using the three different mean velocity profiles in Fig.~\ref{fig:Fukagatakz0}a.  In other words, the compliant wall properties are identical for all three panels, but the mean profile used in the resolvent operator varies.  Visually, the mean velocity profile over the smooth wall (dashed line) is not substantially different from the mean profile observed in the small domain compliant wall DNS (fine gray line) that led to an $8.3\%$ reduction in drag.  In contrast, the mean velocity profile in the large domain compliant wall DNS (bold gray line), which led to a substantial increase in drag, is more rounded and exhibits a much sharper gradient close to the wall.  Consistent with these observations, the predicted changes in singular values over the compliant wall are not significantly different when either the smooth wall (Fig.~\ref{fig:Fukagatakz0}b) or small domain $U(y)$ (Fig.~\ref{fig:Fukagatakz0}c) is used in the Eq.~\ref{eqLinearOperator}.  This lack of sensitivity to the details of the mean profile suggests that the resolvent framework may be used to generate useful predictions regarding the effects of compliant walls\footnote{Or more broadly, all types of flow control that can be represented via linear boundary conditions, \cite{Luhar2014a}} assuming smooth wall mean velocity profiles.  Of course, such predictions should be treated very much as first approximations that provide qualitative insight into the effects of a given compliant wall.  As illustrated by Fig.~\ref{fig:Fukagatakz0}d, the changes in predicted mode amplification are more substantial when the large domain compliant wall $U(y)$ is used instead of the smooth wall or small domain mean profiles.

In many ways, the wall characteristics discussed above (\eg constant, low resonant wave speed) are similar to the more traditional tension-only wall considered in Fig.~\ref{fig:KTS}b.  However, there are some important differences in how the two walls affect spanwise constant modes.  Relative to the tension-only wall considered in \S\ref{sec:results-KTS}, the regions of mode amplification and mode suppression appear to be more distributed in spectral space over the anisotropic compliant wall considered in F2008.  For example, there are no sharp transitions in performance across the resonant frequency.  In addition, while the tension-only wall led to substantial further amplification of large-scale slow-moving modes, such modes are suppressed over the F2008 wall (see $\kx < 6$ and $c/2U_B < 0.3$ in Fig.~\ref{fig:Fukagatakz0}b,c).  We suggest that this mode suppression may be due to the negative Reynolds stress constraint imposed by the kinematic boundary conditions, though a detailed description of the exact mechanism is outside of the scope of the present effort.

Importantly, Fig.~\ref{fig:Fukagatakz0}c shows why the F2008 compliant wall only led to drag reduction in the small domain of length $3h$.  Specifically, the resolvent framework predicts that modes with streamwise wavenumbers smaller than $\kx \approx 2\pi/3$ (vertical gray line) and $c/2U_B > 0.3$ are further amplified over the compliant wall.  This wavenumber cutoff represents structures with streamwise wavelength greater than $3h$, which can only appear in the larger domain of length $6h$.  Of course, there are some regions of increased amplification for $\kx \ge 2\pi/3$ as well.  For instance, amplification increases nearly 20-fold for modes with $\kx = 10$ and $c/2U_B \approx 0.22$ (point e in Fig.~\ref{fig:Fukagatakz0}c) and by a more limited $25\%$ for modes with $\kx = 2\pi/3$ and $c/2U_B \approx 0.35$ (point g in Fig.~\ref{fig:Fukagatakz0}c).  However, as shown in Fig.~\ref{fig:Fukagatakz0}e, the highly-amplified shorter mode is characterized by strong negative (beneficial) Reynolds stress at the wall and only a limited region of positive (detrimental) Reynolds stress for $y<0.1$.  In any case, any detrimental effects associated with these modes that are further amplified is likely offset by the beneficial impact associated with mode suppression across the rest of spectral space.

Figure~\ref{fig:Fukagatakz0}d indicates that structures with $\kx < 2\pi/3$ are amplified even more when the large domain mean velocity profile is used in the resolvent operator.  For instance, the domain-spanning large scale mode with $\kx = 2\pi/6$ and $c/2U_B = 0.42$ (point h in Fig.~\ref{fig:Fukagatakz0}d) experiences a near $300\%$ increase in amplification over the compliant wall.  Further, unlike the other highly amplified modes shown in Fig.~\ref{fig:Fukagatakz0}e-g, which contribute substantial negative Reynolds stresses at the wall, Fig.~\ref{fig:Fukagatakz0}h shows that this domain-spanning mode primarily contributes positive Reynolds stress in the region $0<y<0.3$.  The blue shading at the wall, which denotes negative Reynold stresses, is much less pronounced and there are substantial regions of positive Reynolds stress (red shading) in the fluid domain.

Thus, the larger domain may create a feedback loop whereby two-dimensional wavelike structures with length scales greater than $3h$ arise and generate significant additional detrimental Reynolds stress.  In addition to transferring energy from the mean flow to the turbulence, this Reynolds stress also modifies the mean velocity profile such that the wavelike modes are amplified even further.  This reinforcement could explain the substantial increase in drag observed in the large domain DNS.  However, bear in mind that additional effects such as separation and secondary circulation over the large-amplitude wavelike motion of the compliant wall may also play a role in enhancing drag.  Unfortunately, such effects cannot be captured in the resolvent framework without explicit treatment of nonlinear effects in the forcing $\fk$ (Eq.~\ref{eqNSE}) or boundary conditions.  Specifically, \textit{a priori} prediction of changes to the mean profile and prediction of secondary circulations requires an explicit treatment of the nonlinear forcing terms, while accounting for the effects of large wall deflection and separation requires nonlinear boundary conditions.

Despite these limitations, the results presented in this section indicate that, at the very least, the resolvent formulation may be used as a first-order test of material properties prior to testing in more computationally intensive simulations.  While increasing the domain size in DNS carries a heavy computational penalty, extending the resolvent analysis to lower wavenumbers is inexpensive, especially when coupled with additional physical insight (\eg focus on 2D modes) to limit the region of spectral space to be explored.  As an example, all the results shown in Fig.~\ref{fig:Fukagatakz0} were computed in less than one hour on a single core of a laptop, without any attempt at making the computation efficient.  Further, parameter sweep calculations similar to those shown in Fig.~\ref{fig:Fukagatakz0} are easy to parallelize since each wavenumber-frequency combination is independent.


\section{Conclusion}
The results presented in this paper provide some important design and methodology guidelines for future research on the development of compliant surfaces.  The transitions in performance across the resonant frequency of traditional compliant walls (involving springs, tension, or stiffness) suggest that such walls must be slightly detuned and resonant at frequencies away from the spectral region of interest.  Note that this transition in performance is linked to the phase shift in the relationship between the pressure and wall deflection across the resonant frequency, \ie as $\angle Y$ changes sign, though the phase relationship between the velocity and pressure fields close to the wall also appears to play a role.  Importantly, the amplification-suppression transitions persist at higher $\Ret$ as well, suggesting that it may be possible to generate useful scaling guidelines for compliant walls across all Reynolds numbers.

Although the present study and previous research by Luhar et al. \cite{Luhar2015} employ single wavenumber-frequency combinations as models for VLSMs and the NW-cycle, in reality these structures occupy a region in spectral space.  As such, designing a compliant wall with a sharp transition in performance within this region is unlikely to be effective.  Therefore, instead of optimizing wall performance for a single wavenumber-frequency combination, the optimization must be performed for a range of relevant wavenumbers and frequencies.  In other words, there must be a net decrease in amplification across the entire spectral region of interest.  This will, of course, increase the computational expense associated with the optimization procedure.  However, since the effect of the compliant walls on individual modes (or wavenumber-frequency combinations) can be computed independently, there is significant scope for parallelization.  Keep in mind that the framework in its current form does neglect nonlinear interactions across modes, which serve to generate the forcing terms $\fk = (\ub \cdot \nabla \ub)_\kb$.  While we cannot provide any definitive insight into the nature of these interactions at this point, this is an area of active research for the authors.

The results presented in \S\ref{sec:results-KTS} show that, in general, compliant walls must minimize the susceptibility to spanwise-constant structures to be effective.  This is likely to be difficult given that most natural materials tend to act as low-pass filters.  That is, the effective spring constant generally decreases with decreasing $\kz$.  One potential solution is to employ walls that are in spanwise compression, $C_{tz} < 0$, which would lead to a larger effective spring constant for $\kz = 0$.  Other possibilities include periodic spanwise breaks in the compliant material to disperse the spanwise-constant structures, or the use of mechanical metamaterials which ensure that the curvatures in the streamwise and spanwise directions are coupled, \ie such that surface waves with $\kz = 0$ and $\kx \neq 0$ cannot be generated.

The results presented in \S\ref{sec:results-F2008} suggest that clever anisotropic compliant walls, similar to those proposed by Fukagata et al. \cite{Fukagata2008}, may be used to suppress substantial regions of spectral space.  However, the two-dimensional structures discussed above still play a vital role in dictating overall performance, with larger-scale structures being particularly susceptible to further amplification.  Thus, any simulation-based design of compliant walls requires the use of large computational domains, which is likely to impose severe restrictions on the Reynolds number or the extent of the parameter space that can be explored.

Resolvent analysis provides a computationally-inexpensive alternative that can be used to test and optimize wall properties prior to more detailed evaluation in DNS.  One of the key limitations of the resolvent formulation in its present form is the requirement of a mean velocity profile $U(y)$, which is unknown \textit{a priori} over compliant walls.  However, the results presented in this paper indicate that small changes in the mean profile (\eg associated with a $<10\%$ change in drag) do not substantially alter predictions.  As a result, smooth-wall mean profiles may still be used to generate useful initial predictions.

\section{Acknowledgments}
The authors gratefully acknowledge financial support from AFOSR grant FA9550-12-1-0469 (Program Manager: Doug Smith) and AFOSR/EOARD grant FA9550-14-1-0042 (Program Manager: Russ Cummings).  The authors also thank Professor Koji Fukagata for generously sharing previous DNS results.


\bibliographystyle{tJOT}
\bibliography{2016-Luhar-Compliant}

\begin{thebibliography}{26}
\providecommand{\natexlab}[1]{#1}

\bibitem[1]{Kramer1961}
M.O. Kramer, {\itshape The dolphin's secret}, Journal of the American Society
  for Naval Engineers 73 (1961), pp. 103--108.

\bibitem[2]{Bushnell1977}
D.M. Bushnell, J.N. Hefner, and R.L. Ash, {\itshape Effect of compliant wall
  motion on turbulent boundary layers}, Physics of Fluids 20 (1977), pp.
  S31--S48.

\bibitem[3]{GadelHak1984}
M. {Gad-el-Hak}, R.F. Blackwelder, and J.J. Riley, {\itshape On the interaction
  of compliant coatings with boundary-layer flows}, Journal of Fluid Mechanics
  140 (1984), pp. 257--280.

\bibitem[4]{Lee1993}
T. Lee, M. Fisher, and W.H. Schwarz, {\itshape Investigation of the stable
  interaction of a passive compliant surface with a turbulent boundary layer},
  Journal of Fluid Mechanics 257 (1993), pp. 373--401.

\bibitem[5]{Choi1997}
K.S. Choi, X. Yang, B.R. Clayton, E.J. Glover, M. Atlar, B.N. Semenov, and V.M.
  Kulik, {\itshape Turbulent drag reduction using compliant surfaces},
  Proceedings of the Royal Society A-Mathematical Physical and Engineering
  Sciences 453 (1997), pp. 2229--2240.

\bibitem[6]{Zhang2015}
C. Zhang, R. Miorini, and J. Katz, {\itshape Integrating Mach--Zehnder
  interferometry with TPIV to measure the time-resolved deformation of a
  compliant wall along with the 3D velocity field in a turbulent channel flow},
  Experiments in Fluids 56 (2015), pp. 1--22.

\bibitem[7]{Endo2002}
T. Endo, and R. Himeno, {\itshape Direct numerical simulation of turbulent flow
  over a compliant surface}, Journal of Turbulence 3 (2002), pp. 1--10.

\bibitem[8]{Xu2003}
S. Xu, D. Rempfer, and J. Lumley, {\itshape Turbulence over a compliant
  surface: numerical simulation and analysis}, Journal of Fluid Mechanics 478
  (2003), pp. 11--34.

\bibitem[9]{Fukagata2008}
K. Fukagata, S. Kern, P. Chatelain, P. Koumoutsakos, and N. Kasagi, {\itshape
  Evolutionary optimization of an anisotropic compliant surface for turbulent
  friction drag reduction}, Journal of Turbulence 9 (2008), pp. 1--17.

\bibitem[10]{Kim2014}
E. Kim, and H. Choi, {\itshape Space-time characteristics of a compliant wall
  in a turbulent channel flow}, Journal of Fluid Mechanics 756 (2014), pp.
  30--53.

\bibitem[11]{Luhar2015}
M. Luhar, A.S. Sharma, and B.J. McKeon, {\itshape A framework for studying the
  effect of compliant surfaces on wall turbulence}, Journal of Fluid Mechanics
  768 (2015), pp. 415--441.

\bibitem[12]{McKeon2010}
B.J. McKeon, and A.S. Sharma, {\itshape A critical-layer framework for
  turbulent pipe flow}, Journal of Fluid Mechanics 658 (2010), pp. 336--382.

\bibitem[13]{Moarref2013}
R. Moarref, A.S. Sharma, J.A. Tropp, and B.J. McKeon, {\itshape Model-based
  scaling of the streamwise energy density in high-Reynolds-number turbulent
  channels}, Journal of Fluid Mechanics 734 (2013), pp. 275--316.

\bibitem[14]{Sharma2013}
A.S. Sharma, and B.J. McKeon, {\itshape On coherent structure in wall
  turbulence}, Journal of Fluid Mechanics 728 (2013), pp. 196--238.

\bibitem[15]{Luhar2014b}
M. Luhar, A.S. Sharma, and B.J. McKeon, {\itshape On the structure and origin
  of pressure fluctuations in wall turbulence: predictions based on the
  resolvent analysis}, Journal of Fluid Mechanics 751 (2014), pp. 38--70.

\bibitem[16]{Luhar2014a}
M. Luhar, A.S. Sharma, and B.J. McKeon, {\itshape Opposition control within the
  resolvent analysis framework}, Journal of Fluid Mechanics 749 (2014), pp.
  597--626.

\bibitem[17]{Weideman2000}
J.A.C. Weideman, and S.C. Reddy, {\itshape A {MATLAB} differentiation matrix
  suite}, ACM Transactions on Mathematical Software 26 (2000), pp. 465--519.

\bibitem[18]{McKeon2013}
B.J. McKeon, I. Jacobi, and A.S. Sharma, {\itshape Experimental manipulation of
  wall turbulence: a systems approach}, Physics of Fluids 25 (2013), p. 031301.

\bibitem[19]{Reynolds1967}
W.C. Reynolds, and W.G. Tiederman, {\itshape Stability of turbulent channel
  flow with application to {M}alkus's theory}, Journal of Fluid Mechanics 27
  (1967), pp. 253--272.

\bibitem[20]{Duvvuri2015triadic}
S. Duvvuri, and B.J. McKeon, {\itshape Triadic scale interactions in a
  turbulent boundary layer}, Journal of Fluid Mechanics 767 (2015), p.~R4.

\bibitem[21]{Fukagata2002}
K. Fukagata, K. Iwamoto, and N. Kasagi, {\itshape Contribution of {R}eynolds
  stress distribution to the skin friction in wall-bounded flows}, Physics of
  Fluids 14 (2002), pp. 73--76.

\bibitem[22]{Marusic2010}
I. Marusic, R. Mathis, and N. Hutchins, {\itshape Predictive Model for
  Wall-Bounded Turbulent Flow}, Science 329 (2010), pp. 193--196.

\bibitem[23]{Carpenter1990anisotropic}
P.W. Carpenter, and P.J. Morris, {\itshape The effect of anisotropic wall
  compliance on boundary-layer stability and transition}, Journal of Fluid
  Mechanics 218 (1990), pp. 171--223.

\bibitem[24]{Grosskreutz1971}
R. Grosskreutz {\itshape Wechselwirkungen zwischen turbulenten Grenzschichten
  und weichen W{\"a}nden},    Selbstverlag Max-Planck-Institut f{\"u}r
  Str{\"o}mungsforschung und der Aerodynamische Versuchsanstalt, 1971.

\bibitem[25]{CUEDMechanics}
 {Cambridge University Engineering Department}, Mechanics Data Book;  (2000), .

\bibitem[26]{GadelHak1986}
M. {Gad-el-Hak}, {\itshape The response of elastic and viscoelastic surfaces to
  a turbulent boundary layer}, Journal of Applied Mechanics 53 (1986), pp.
  206--212.

\end{thebibliography}
\end{document}